# Intermediate Phases, structural variance and network demixing in chalcogenides: the unusual case of group V sulfides

P. Boolchand, Ping Chen and U. Vempati.*
Department of Electrical and Computer Engineering
University of Cincinnati, Cincinnati
OH 45221-0030

Abstract

We review Intermediate Phases (IPs) in chalcogenide glasses and provide a structural interpretation of these phases. In binary group IV selenides, IPs reside in the $2.40 < r < 2.54$ range, and in binary group V selenides they shift to a lower $r$, in the $2.29 < r < 2.40$ range. Here $r$ represents the mean coordination number of glasses. In ternary alloys containing equal proportions of group IV and V selenides, IPs are wider and encompass ranges of respective binary glasses. These data suggest that the local structural variance contributing to IP widths largely derives from *four isostatic* local structures of varying connectivity $r$; two include group V based quasi-tetrahedral ($r = 2.29$) and pyramidal ($r = 2.40$) units, and the other two are group IV based corner-sharing ($r = 2.40$) and edge-sharing ($r = 2.67$) tetrahedral units. Remarkably, binary group V (P, As) sulfides exhibit IPs that are shifted to even a lower $r$ than their selenide counterparts; a result that we trace to excess $S_n$ chains either partially (As-S) or completely (P-S) demixing from network backbone, in contrast to excess $Se_n$ chains forming part of the backbone in corresponding selenide glasses. In ternary chalcogenides of Ge with the group V elements (As, P), IPs of the sulfides are similar to their selenide counterparts, suggesting that presence of Ge serves to reign in the excess $S_n$ chain fragments back in the backbone as in their selenide counterparts.

1. **Three types of glass transitions**

Glasses are intrinsically non-equilibrium solids and their physical properties generally evolve over long times, i.e., these solids slowly age. The aging of glasses is itself a fascinating subject and has been debated since the early work of Kohlrausch [1-3]. There is now evidence to suggest that the stretched exponential relaxation observed in glasses may well be characterized by specific exponents, which are determined[2] largely by the nature ( of long or short range) of forces that control how traps or defects diffuse as networks relax. For a long time it was widely believed that glass transitions are also hysteretic and age[3] as observed in a traditional Differential Scanning Calorimetry. There are new findings showing that under select conditions [4-6] aging of glasses may not occur.

New insights into the nature of the glass transition [4, 7] have now emerged using modulated- DSC (m-DSC). A significant advantage of m-DSC over traditional DSC is that the method permits deconvoluting the *total* heat flow into a *reversing* heat flow term (which captures the local equilibrium specific heat ) and a *non-reversing* heat flow term (which captures non-equilibrium effects of the changing structure). The thermally reversing heat flow term usually reveals a rounded step-like jump. One defines the glass transition temperature, $T_g$, from the inflexion point of the step, and the specific heat jump, $\Delta C_p$ in going from the glass to the liquid state from the height of the step. On the other hand the *non-reversing* heat flow term usually shows a Gaussian like peak as a precursor to $T_g$, and the integrated area under the Gaussian lineshape, provides the *non-reversing enthalpy* ($\Delta H_{nr}$) of $T_g$. Experiments on wide variety of glasses reveal [8-10] that the $\Delta H_{nr}$ term depends on factors such as sample purity, sample homogeneity and sample aging. Kinetic factors such as scan rate and modulation rate used to record a scan also influence $\Delta H_{nr}$ and $T_g$, although their influence can be corrected by a judicious choice of procedure in these experiments[11].

Experimental data on covalently bonded glasses examined systematically as a function of their network connectivity (or mean coordination number *r*) show that there are, in general, *three types*[4, 7] of glassy networks formed in freezing as characterized by their elasticity. The first type is the elastically *flexible* networks which form at low connectivity (*r* ~ 2), such as a Se glass consisting of a chain in which each atom has 2 neighbors. Such networks display a non-reversing enthalpy ($\Delta H_{nr}$) of $T_g$ that is usually narrow (~15°C) and symmetric in temperature and, which slowly ages with waiting time as a stretched exponential[2]. The *second* type of glass network is elastically *rigid but stress-free*. These networks form at intermediate connectivity (*r* ~ 2.4); the endotherm has a *vanishing* $\Delta H_{nr}$ term that *shows little or no aging*. The *third* type of glass network is elastically rigid but also stressed. They occur at high connectivity ( *r* ~ 3), and show a $\Delta H_{nr}$ profile that is broad and asymmetric with a high-T tail, and age with waiting time. In this category are chalcogenide glasses with covalently bonded alloys of group IV (tathogen, *r* = 4) and group V (pnictide, *r* = 3 ) elements with group VI (chalcogen, *r* = 2) elements. They usually bond in conformity with the 8-N rule, thus making possible an estimate of their connectivity from their chemical stoichiometry alone provided that the resulting components do not segregate[12] or demix on a nano- or microscale[13]. These systems are particularly attractive because they form bulk glasses over wide range in connectivity $r$, permitting calorimetric[14], dielectric[15-17], optical[18-21] and electrical[22] measurements to be performed to establish the global behavior. Remarkably, physical properties of glasses can sometimes change with composition sharply[23, 24] rather than slowly[12, 25, 26]. Thus it can be risky to infer the global behavior by merely investigating stoichiometric glass compositions alone.

Recently, we have examined ionically bonded[24, 27] (alkali- germanates and -silicates ) and fast-ion conducting[28] ( solid electrolyte) glasses in m-DSC experiments and have confirmed the three elastic phases mentioned above. These new findings underscore the *generic* nature of the *three types of glass phases*. Indeed, a simple measurement of the glass transition endotherm in an m-DSC experiment makes it now feasible to identify whether a glass sample possesses a *flexible* network, or a *rigid but stress-free* one, or a *rigid but stressed* one.

2. **Self-organization and reversibility windows in chalcogenide glasses**

As mentioned earlier, in a wide variety of systems the three types of glass transitions mentioned above occur sequentially with composition as connectivity of their networks is systematically increased. The most unexpected finding[25, 29-31] is the existence of the *second* type of glass structure, which spans a small range of compositions with rather sharply defined edges in some cases. These compositional windows are known as *reversibility windows* (RWs) since $T_g$s become almost completely thermally reversing ($\Delta H_{nr} \sim 0$). These windows represent calorimetric signatures of different vibrational regimes in networks as we discuss next.

The inspiration to look for these elastically special network glasses came in the early 1980's from the pioneering work of Phillips[32] and Thorpe[33]. They predicted[32-34] the existence of a *solitary* elastic phase transition in a covalent amorphous network from an elastically flexible phase to a stressed-rigid phase when its connectivity increases to $r$ =2.40. Lamb-Mossbauer factors in $^{119}$Sn Mossbauer spectroscopy of binary $Ge_xSe_{1-x}$ glasses confirmed[31, 35, 36] the existence of a vibrational threshold behavior in them. Starting in 1996, detailed Raman scattering experiments on two group IV selenide glasses (Si-Se and Ge-Se) revealed the existence of *two*[31, 37] elastic phase transitions and not the isolated one predicted. Numerical simulations on self-organized networks subsequently showed[38] that the first transition at low $r$ (= $r_1$) was between *floppy* (or *flexible*) and *rigid* phases, while the second transition at higher $r$ (= $r_2$ ) was to a *stressed* network. The intervening region, ($r_1$ < r < $r_2$), also called the Intermediate Phase (IP), separates the *flexible* phase from the *stressed-rigid* one . Subsequently, experiments on several other glass systems[5, 26, 39-43] including ternary chalcogenides, confirmed that IPs observed in vibrational spectroscopy coincide with RWs observed in calorimetric measurements [31, 37]. These optical and thermal measurements are but two fingerprints of the IP as being a remarkable new kind of disordered solid[6]. Glassy networks in this phase are rigid but stress-free. Their quasi-equilibrium state is much like a crystalline solid, with nearly absent aging of structure.

There have been several attempts in recent years to simulate IPs in amorphous networks numerically. For example, the vibrational behavior of 3D amorphous Si networks[38, 44] provided suggestive evidence of an IP existing in a narrow range of connectivity, 2.376 < r < 2.392. IPs have also been observed in 2D triangular networks and their nature exhaustively explored by equilibrating networks using activation relaxation techniques[44]. Although chemically not realistic, these models serve to confirm the existence of IPs in numerical simulations, bringing theory and experiments closer together. The rigidity and stress phase boundaries of more realistic analytic models consisting of small networks formed by agglomerating corner – sharing (CS) and edge-sharing (ES) tetrahedra have also displayed an IP whose width appears to be controlled by the fraction of ES to CS tetrahedra. These analytical calculations[45, 46] on binary Si-Se and Ge-Se glasses predict IP widths that are much closer to experiments. Mauro and Varshneya[47, 48] modeled binary $Ge_xSe_{1-x}$ glasses using empirical potentials, and were able to provide evidence of a rigidity transition near *r* ~2.4. Numerical simulations of the IP in real glasses pose formidable challenges not the least of which is constructing large and space filling structural models with appropriate local structures. This is an area of active interest in the field. [49, 50]

The aim of the present review is to focus on IPs of the group V- sulfides and selenides. The IP in As-Se[26] and P-Se[41] binaries were reported a few years ago. Now, IPs in binary As-S[9], and P-S[51] glasses have become available. These new results on group V-chalcogenides along with earlier ones on group IV ones provide a platform to discuss trends in IPs in general, and we will address here such questions as what aspects of local and intermediate range structure of glasses control the width and centroid of the IP in these covalently bonded solids?

In section 3, we discuss trends in $T_g$ and molar volumes, which provide important clues on glass structure.

In Section 4, we give an overview of observed RWs in group V- sulfides, and in section 5, Raman scattering results on these systems. These data then permit a discussion of the IPs and their connection to glass structure in chalcogenides in Section 6. A summary of our findings appears in the conclusions.

## 3. Compositional trends of $T_g$

*3.1. Variation of Glass Transition temperatures-role of network connectivity.*

In the past 8 years reliable data on the variation of $T_g$ as a function of glass composition in binary and ternary chalcogenide glasses has evolved using m-DSC. What underlying physics resides in these findings? Can we connect these trends to aspect of glass structure? Here we will show that the connectivity of the underlying networks controls in a crucial fashion the observed variations of $T_g$. Many of these ideas have been made possible by Stochastic Agglomeration Theory[52] and Lindeman melting criteria[36].

*3.1.1. Group IV selenides*

Perhaps the simplest cases to consider are those of the $Si_xSe_{1-x}$[30] and $Ge_xSe_{1-x}$[53] binaries where rather complete trends in $T_g(x)$ are now available from m-DSC measurements (Figure.1). One finds that the variation at low x ( < 10%) for both binary glass systems can be described by a slope $dT_g/dx \sim 4.4°$ C/at.% of Si or Ge. In these binary glasses, the SAT predicted[52] slope is given by

$$dT_g/dx^{SAT} = T_o / \ln (r_{Si\ or\ Ge}/r_{Se}) \qquad (1)$$

and equals 4.5°C/at.% Si or Ge. Here $T_o$ is the glass transition temperature of the base glass of Se (= 40°C) and $r_{Si\ or\ Ge}$ and $r_{Se}$ represent the coordination numbers of the group IV atoms and Se atoms and are taken respectively as 4 and 2. The observed slope is in excellent agreement[53] with experiments at low x ( < 10%) where the cross-linking of $Se_n$ chains by the group IV additives proceeds stochastically. For group IV additives there is wide recognition that $r = 4$. A perusal of the data of Figure 1 also shows that while $T_g(x)$ increase with x for both binary systems, in the 10% < x < 20% range, the rate of increase of $T_g$ is lower in the Si-Se binary than in the Ge-Se binary. In this range of composition it is also known from Raman scattering that the concentration of edge-sharing (ES) tetrahedra in the Si-Se[37] binary exceeds that in the Ge-Se binary[25, 37]. There is, thus, a greater number of ways in which CS tetrahedra can link with ES ones in the Si-Se binary than in the Ge-Se binary. The increased entropy of bonding configurations, it is thought[52] leads to slower increase of $T_g$ in Si-Se glasses than in Ge-Se ones, even though chemical bond strength considerations would suggest otherwise. The Si-Se bond strength[54] (51.4 kcal/mole) slightly exceeds that of Ge-Se bonds (49.1 kcal/mole). At x > 33.33%, one finds that $T_g$s of Si-Se glasses continue to increase while those of Ge-Se glasses decreases with increasing x. The threshold behavior of $T_g$ in binary Ge-Se glasses is the result of nanoscale phase separation[55], with Ge-Ge bonds segregating from the network backbone. On the other hand, in Si-Se glasses Si-Si bonds also form at x > 33.3%, but they do so as part of the network, and one finds $T_g$s to continue to increase with x. These data underscore the fact that network connectivity considerations are paramount, and these overwhelm chemical bond strength considerations in determining compositional trends of $T_g$ in glasses. We shall revisit these ideas again in connection with the group V selenides.

*3.1.2 Group V selenides*

Figure 2 summarizes compositional trends in $T_g$ for binary P-Se and As-Se glasses. One finds that the slope at low x ( < 10%) is 4.1° C/at.% of As and 3.6°C/at.% of P for the two sets of data. The SAT based prediction[52] of the slope, $dT_g/dx$, gives

$$dT_g/dx^{SAT} = T_o/\ln(r_{As}/r_{Se}) \qquad (2)$$

and yields a value of 7.7°C/at.% As or P, if one takes the coordination number *r* of the group V elements as 3 and $T_o$, the $T_g$ of Se as 40°C. Thus, the SAT prediction of the slope, $dT_g/dx$, for group V Selenides disagrees with the observed slope $dT_g/dx$ rather noticeably if As and P coordination numbers are taken to be 3-fold. It is widely believed that As takes on a 3-fold coordinated local structure in a pyramidal (PYR) As(Se$_{1/2}$)$_3$ units in Se-rich ( x < 2/5) binary As$_x$Se$_{1-x}$ glasses. Several years ago, Georgiev et al.[26] suggested that the lower slope observed in As$_x$Se$_{1-x}$ glasses probably results from the presence of a finite concentration of As atoms present in a QT local structure (Se=As(Se$_{1/2}$)$_3$) with a coordination number, $r_{As}$ = 4. In such a local structure As has 3 bridging and a non-bridging Se near neighbor.

For binary P$_x$Se$_{1-x}$ glasses (Figure 2) the observed slope, $dT_g/dx$, at low x ( < 0.15) of 3.6°C/at. %P, is again found to be lower than the SAT predicted value of 7.7°C/at.% of P. The discrepancy, it has been suggested[56], can be reconciled if one assumes P to be also 4-fold coordinated in addition to 3-fold. In the case of P-Se glasses, both P$^{31}$ NMR and Raman scattering provide unambiguous evidence[56, 57] of QT units, PYR units and ethylenelike P$_2$Se$_2$ (ETY) units. NMR results[57] show the ratio of 4-fold to 3-fold coordinated P, $P_4/P_3$, to decrease almost linearly from a value of 1 at x ~0 to a value of 0 as x increases to 2/5. These data suggest that the absolute fraction of P-atoms that are 4-fold to 3-fold coordinated then varies as $xP_4/P_3$, and maximizes near x ~ 0.25. Taken together, these results on binary As-Se and P-Se bulk glasses reveal a *commonality* in which the group V additives modify the chain-structure of the base Se glass by `acquiring both a 3-fold and a 4-fold coordination.

A perusal of the compositional trends of $T_g$ in the As- and P- selenides at higher x ( > 10%) reveal (Figure 2) other surprises that can be traced to aspects of network connectivity related to local structures. For binary P-Se glasses, one observes almost a plateau in $T_g(x)$ in the 20% < x < 40% range even though chemically stronger P-Se (49.75 kcal/mole) and P-P ( 51.3 kcal/mole) are being formed at the expense of Se-Se bonds (44 kcal/mole) as the P-content of glasses increases. Experiments reveal that there are three types of P-centered local structures formed in these glasses that crosslink $Se_n$ chains, and these include pyramidal (PYR) , quasi-tetrahedral (QT) and ethylene like (ETY) units. The multiplicity of these local structures increases the entropy of bonding configurations, and the slope, $dT_g/dx$, depends inversely on the entropy term.

In the 40% < x < 50% range, $^{31}$P NMR[56] and Raman scattering[56] experiments reveal that predominantly only ETY units occur in glass structure. The aspect of local structure reduces the number of ways in which these units can connect and leads the slope, $dT_g/dx$, to increase remarkably. In As-Se glasses only two As based local structures ( QT and PYR) occur in the 15% < x < 35% range, and the slope ($dT_g/dx$) is found to be larger than in P-Se glasses, where three local structures exist as mentioned above. These data again suggest that network connectivity considerations overwhelm chemical bond strength ones when compositional trends in $T_g$ are considered. Indeed, if latter considerations alone were to play a part, and if $P_2Se_3$ glass like $As_2Se_3$ glass were to be composed of PYR units alone, one would have expected the $T_g$ of $P_2Se_3$ glass to exceed that of $As_2Se_3$ glass by 19%, given that P-Se bonds ( 49.75 kcal/mole) are chemically stronger than As-Se bonds ( 41.73 kcal/mole) by 19%.

At x > 50%, $T_g(x)$ of both P-Se and As-Se glasses steadily decrease with increasing x largely because both glass systems demix on a nanoscale, with $P_4Se_3$ molecules in the former and $As_4Se_4$ related monomers in the latter segregating from the backbone. The circumstance is analogous to the maximum of $T_g$ observed in Ge-Se glasses near the chemical threshold ( x ~33.3%) as discussed above.

*3.2. Variation of glass transition temperatures-role of chemical bonding*

For networks possessing the same connectivity one expects $T_g$s to scale with chemical bond strengths[58]. An illustrative example is the $T_g$ of $GeS_2$[8] (508°C) which is 13.2% higher than $T_g$ of $GeSe_2$ glass[53] (416°C). The Pauling single Ge-S chemical bond strength[54] (55.52 kcal/mole) exceeds that of a Ge-Se (49.08 kcal/mole) bond by 13.1%, and provides a quantitative rationale for the increased $T_g$ of the sulfide glass compared to the selenide glass. The analogy appears to hold in other systems including the $P_xGe_xX_{1-2x}$ ternaries [51, 59] with X = Se or S, as illustrated in Figure 3. A perusal of data shows a close parallel in $T_g$s of the two ternaries in the 10% < x < 18% range. In this range of composition the observed scaling of $T_g$s quantitatively correlates with the higher chemical bond strengths in the sulfide glasses (Ge-S, P-S) compared to the selenides ( Ge-Se, P-Se). These data suggest that the underlying network connectivity must be similar. In particular there must be little or no evidence of demixing of backbones in this range. In the ternary sulfides[51], that situation changes drastically once x < 10% as the more stable $S_8$ rings demix from the backbone, and one observes a linear reduction of $T_g$ which extrapolates to a value of about -50°C at x = 0, which we assign to the $T_g$ of a $S_8$ ring glass. The steady demixing of $S_8$ rings is corroborated by the appearance of the sulfur polymerization transition, $T_\lambda$ transition, in calorimetry[9, 60] and sharp modes in Raman scattering[9, 60].

The $T_g$ (x) variation in binary P-Se and P-S glasses reveal other features of interest. In the narrow range, 20% < x < 23%, the observed ratio of their glass transition temperatures, $T_g(P-S)/T_g(P-Se)$, is close to 1.08, which is somewhat below the expected chemical bond strength scaling ratio of $E_b$ (P-S)/$E_b$(P-Se) = (54.78 kcal/mole)/(49.75 kcal/mole) of 1.10 . At x < 16% and also at x > 23% the observed $T_g(P-S)/T_g(P-Se)$ scaling ratio decreases monotonically . These trends show that it is only in the narrow interval, 20% < x < 23% , that the P- sulfide and selenide glasses have some similarity of network structure. However, at P compositions x < 12% and at x > 23%, there is no semblance of any similarity in glass structure. And we shall see later, $S_8$ rings demix at low x ( <12%) while $P_4S_{10}$ molecules do so at high x ( > 23%) in the P-S binary. In retrospect, it is somewhat remarkable that by alloying Ge in both the binary glass systems , P-S, and As-S, one is able to restore a striking similarity of glass structure over a wide range of compositions that lead to chemical bond strength scaling of $T_g$s in the Ge-P-X with X = S or Se ternaries (Figure 3).

If the molecular structure of binary $As_xX_{1-x}$ glasses with X = Se or S, were to possess fully polymerized networks of closely similar local and intermediate range structures, one would have expected their $T_g$s to vary in a parallel fashion allowing for a scaling of these temperatures by their chemical bond strengths. In Figure 5 we plot the measured $T_g$s of As-Se glasses[26] along with a 13.2% scaled variation of these $T_g$s ( which smoothly increases to 15.7% at x = 0) as shown by the dotted line. The measured $T_g$s of As-S glasses, from recent work of Chen et al.[9] shown in Figure 4, are found to deviate *qualitatively* from the predicted scaling of $T_g$s. The 13.2% scaling reflects the higher single bond strength[54] of As-S bonds (47.25 kcal/mole) compared to As-Se (41.73 kcal/mole) ones, while the 15.7% scaling represents the higher bond strength of S-S bonds (50.9 kcal/mole) compared to Se-Se bonds ( 44.0 kcal/mole)[54]. The absence of $T_g$ scaling must imply that the connectivity of As-S glass structure must be lower than of As-Se glass structure. Binary $As_xS_{1-x}$ glasses can only be regarded as being partially polymerized if $As_xSe_{1-x}$ glasses represent examples of nearly fully polymerized networks. In the 25% < x < 38% range, where large scale demixing effects[9, 26] are absent in both systems, the conspicuous absence of scaling of $T_g$s underscores that As-S glasses must form networks that are not fully connected. In particular, the molecular structure of the stoichiometric $As_2S_3$ glass having a $T_g$ that is nearly the same as that of $As_2Se_3$ glass, strongly suggests that it *cannot* be an example of a fully polymerized network[60] if $As_2Se_3$ glass is an example of one. The glass structure picture evolving from these thermal data is corroborated by network packing considerations as we illustrate next.

*3.3. Molar Volumes of As-S and As-Se glasses compared*

It is instructive to compare molar volumes of c-$As_2Se_3$[61] with c-$As_2S_3$[62] since their crystal structures, and therefore network connectivity are identical. The 15% lower molar volume of c-$As_2S_3$ compared to c-$As_2Se_3$ (Figure 6) can be viewed as bond length rescaling of the unit cell due to the replacement of Se by undersized S in the orpiment structure, an entirely "chalcogen size effect". On the other hand, in glasses in the 25% < x < 40% range, where network formation is thought to occur, molar volumes of $As_xS_{1-x}$ glasses are only 9% lower than of $As_xSe_{1-x}$ glasses (Figure 6). If selenide glasses are examples of fully polymerized structures, then these data suggest that the sulfide glasses possess networks that can only be partially polymerized, i.e., they must contain substantial *free* volume. Such free volume most likely can comes from some excess $S_n$ chain fragments decoupling from the backbone by forming non-bridging S, i.e., forming van der Waals bonds rather than covalent ones alone. In summary, both chemical trends in $T_g$ and molar volumes, strongly suggest that As-S glasses are examples of *partially polymerized* networks.

## 4. Reversibility windows in group V-chalcogenides

*4.1. Binary As-Se glasses*

The RW in As-Se glasses was reported by Georgiev et al[26] to lie in the 29% < x < 37% range, or 2.29 < *r* < 2.37 range of connectivity. The $\Delta H_{nr}$ term increases as x increases to about 50%, and then decreases thereafter to nearly vanish as x approaches to 60% (Figure 7). At x > 40%, de-mixing of the glassy networks takes place as As-rich clusters decouple from the backbone and lead to a maximum in $T_g$ and a leveling off in the $\Delta H_{nr}$ term as x exceeds 50%. The vanishing of the $\Delta H_{nr}$ term as x = 60% is signature of segregation of the glass structure resulting in substantial loss of network backbone. Not surprisingly, a glass at x = 60% has been found to be soft and flexible as revealed by presence of low frequency vibrational excitations ( ~5meV) in inelastic neutron scattering experiments of Effey and Cappelletti[63]. These results serve to illustrate an important caveat- estimates of network connectivity (*r*) of a glass composition from its chemical stoichiometry alone will *not* be reliable if the underlying network is demixed or phase separated on a nanoscale[12].

Recently we have re-measured the non-reversing enthalpy of the As-Se samples used in the earlier study of Georgiev et al.[12]. After seven years of aging at room temperature, the non-reversing enthalpy of these glass samples lead to a RW that is not only intact but the window narrows somewhat and sharpens. The elastic phase boundaries defining the IP become *abrupt* as networks reconfigure slowly upon aging to expel stressed bonds from that phase. Calorimetric measurements also suggest[64] the appearance of sub-$T_g$ endotherms upon aging. These results are at variance with another report[65] and will be discussed elsewhere[64].

*4.2. Binary P-Se glasses*

The RW in binary $P_xSe_{1-x}$ glasses was established by Georgiev et al.[41] nearly five years ago, and resides in the 28% < x < 40% range (Figure 8). It translates to a connectivity that spans the 2.29 < *r* < 2.40 range, and is remarkably similar to the one in binary $As_xSe_{1-x}$ glasses. In binary P-Se glasses, evidence for three distinct P-centered local structures is available from Raman scattering and $^{31}$P NMR experiments[56, 57, 66]. These local structures include pyramidal (PYR) $P(Se_{1/2})_3$, quasi-tetrahedra (QT), $Se=P(Se_{1/2})_3$, polymeric ethylenelike (ETY)) $P_2Se_2$ units[41, 56]. A count of bond-stretching and bond-bending constraints reveals that the PYR and QT units are isostatic ($n_c = 3$), while the ETY units are mildly stressed-rigid ($n_c = 3.25$). Since the chemical stoichiometry of these units are $r$ = 2.29 for QT, 2.40 for PYR and 2.50 for ETY, one expects their concentrations to maximize near the chemical compositions, x, of 29%, 40% and 50% respectively. Concentrations of these units have been deduced by $^{31}$P NMR measurements, and the data broadly confirm the predictions (Figure 9). These local structures have characteristic vibrational modes that are resolved in Raman scattering. Figure 10 shows a typical result taken from the work of Georgiev et al. [56], with mode assignments based on (i) first principles cluster calculations and (ii) compositional trends in mode scattering strengths as a function of glass composition. The assignment of the mode near 210 cm$^{-1}$ differs from an earlier one made by Georgiev et al.[56]. In Raman scattering, mode scattering strengths can, in principle, be related to concentration of local structures if mode cross-sections ( matrix element effects) are established. On the other hand, in NMR signal strengths are a direct manifestation of local structure concentrations. Inspite of these details, a perusal of the data shows that general trends in compositional dependence of local structure concentrations deduced from NMR and mode scattering strengths from Raman scattering experiments are in reasonable accord with each other.

*4.3. Binary As-S glasses*

The nature of glass transitions in binary on $As_xS_{1-x}$ glasses has been examined in m-DSC experiments by Chen et al.[9, 67, 68] In the S-rich range (x < 0.25) two endothermic events are observed, one representing $T_g$ ( 60°C) of the backbone and the other near 140°C , the sulfur polymerization ( $T_λ$) transition. The $T_λ$ transition relates to the opening of $S_8$ rings to form chains leading to an enormous increase of melt viscosities. The $T_λ$ transition in orthorhombic sulfur has a non-reversing enthalpy associated with it (Figure 11 bottom panel). In an $As_xS_{1-x}$ glass at x = 15% for example, the $T_λ$ transition has a precursive exothermic event (Figure 11 top panel), suggesting that $S_8$ rings first reconstruct with the backbone releasing heat (exotherm) before an uptake of heat (endotherm) to open and form chains. Note that in the m-DSC scan of the 15% glass sample, the polymorphic α-β transition of elemental S near 118°C is conspicuously absent. We did not observe this transition even in S-richer glasses (x ~ 8%), suggesting that $S_8$ rings do not aggregate to form precipitates of orthorhombic Sulfur in these glasses. Furthermore, we also did not observe presence of orthorhombic S precipitates in S-rich ( x ~8.3%) ternary $Ge_xAs_xS_{1-2x}$ glasses[42], but note that in another study of these ternary glasses, Kincl and Tichy[69] observed these precipitates in their samples even up to x ~15%. During synthesis, we have found necessary to alloy the starting materials for periods exceeding 3 days above the liquidus to achieve homogeneity during synthesis. Sample homogeneity can be ascertained by examining Raman scattering without opening the quartz tube used for synthesis. Returning to the results on binary As-S glasses, the linear decrease of $T_g$ at x < 23% (Figure 5) is precisely the expected result as $S_8$ rings decouple from the backbone as x approaches 0 and weaker inter-ring interactions ( van der Waals) steadily replace the stronger intra-chain ones (covalent ).

The compositional trend in the $\Delta H_{nr}$ term in binary As-S glasses is compared to the one in binary As-Se glasses in Figure 8. These data reveal the RW in As-S glasses to reside in the 22.5% < x < 29.0% range, shifted to lower x in relation to the window in binary As-Se glasses. Molar volumes ( Figure 6), and variations in $T_g$ (Figure 5) in these binary glasses were compared earlier. The structure implications of these data will be commented next as we review Raman scattering results.

*4.4. Binary P-S glasses*

The glass forming tendency in the $P_xS_{1-x}$ binary is restricted[51] to a much narrower range , 5% < x < 25% , of compositions than in corresponding selenides. The P-sulfides tend to segregate into monomers such as $S_8$, $P_4S_{10}$, $P_4S_7$ limiting the range of bulk glass formation. Bulk glasses in the P-S binary were synthesized by handling the starting materials, elemental P and S in a dry ambient and reacting them in evacuated quartz tubings for extended periods.[51] These samples were then characterized by m-DSC and Raman scattering measurements[51]. Trends in $T_g(x)$ and the $\Delta H_{nr}(x)$ in P-S glasses are compared to those in corresponding selenides in Figure 4 and Figure 8 respectively. Even though the range of glass formation in P-S glasses is rather limited, there exists a sharp, deep and narrow reversibility window in the 14% < x < 19% composition range centered in the glass forming range. These data highlight in a rather striking fashion the close relationship between optimization of the glass forming tendency and the RW. We find the RW in P-S glasses to be shifted to significantly lower *r* in relation to P-Se glasses (Figure 8).

We would like to conclude this section with a general comment. In metallic glass systems supercooling is facilitated near eutectics and there is evidence to suggest that bulk glass formation is optimized near these compositions. In chalcogenides, particularly sulfides and selenides of the group IV and group V elements, on the other hand, we see no correlation between RWs where the glass forming tendency is optimized and eutectics. For example, in As-Se binary a eutectic is suggested near 20% of As [70], a composition that resides outside the RW of 28% < x < 38% (Figure 7). In the As-S binary there is no eutectic in the S-rich range, but a RW occurs in the 22% < x < 29% range (Figure 7). This is the exception that proves the rule, viz., eutectics in these good glass forming systems have no bearing on RWs. In P-Se binary a eutectic [70] occurs near 26% of P while the RW occurs in the 28 < x < 40% range (Figure 8). In P-S binary, a eutectic occurs [70] near 7% of P, which lies outside the RW of 14% < x < 19% Figure 8. Fundamentally, optimization of the glass forming tendency in RWs has an *elastic* origin, which leads the configurational entropy between liquid- and glass-structures to be minimal as reflected in a vanishing $\Delta H_{nr}$ term. In these RWs, minuscule changes of structure occur upon arrest of atomic motion upon freezing ( $T < T_g$ ) into the glassy state. These ideas are consistent with our experience that compositions in the RWs form bulk glasses even when melts are cooled slowly by an air quench instead of a water quench.

## 5. Raman scattering and glass structure of group V-sulfides

*5.1. Binary $As_xS_{1-x}$ glasses*

Raman scattering in binary As-S glasses[67, 71] has been examined by several groups. The observed lineshapes shown in Figure 12, taken from the data of Chen et al.[9] are representative of previous results as well. First principles DFT calculations[72] on characteristic clusters place the symmetric and asymmetric vibrations PYR units near 352 and 355 cm$^{-1}$ respectively, and those of QT units near 335 cm$^{-1}$ and 365 cm$^{-1}$ respectively. These calculations also place the symmetric stretch of As=S double bond mode in QT units near 537 cm$^{-1}$. The observed Raman lineshape of a glass sample at x = 15% is deconvoluted as a superposition of several Gaussians. Based on the cluster calculations[72], we propose the following Raman mode assignments (Figure 12); $S_8$ units contribute[73] to a vibrational modes near 150 cm$^{-1}$ , and near 217 cm$^{-1}$, and near 485 cm$^{-1}$, while the broad mode near 430 cm$^{-1}$ represents second order scattering from the intense mode near 217 cm$^{-1}$ mode; $S_n$-chain fragments contribute to a broad mode[73] near 460 cm$^{-1}$ ; PYR units contribute to a broad mode near 370 cm$^{-1}$ ; QT units contribute to modes at 335 cm$^{-1}$ , 375 cm$^{-1}$ and 490 cm$^{-1}$. The mode near 370 cm$^{-1}$ encompasses scattering from both the symmetric and asymmetric mode of PYR units.

The observed lineshapes at other glass compositions can be deconvoluted in a similar manner for As content up to x= 35%, and the normalized mode scattering strengths are summarized in Figure 13. Several trends become apparent from these data. The scattering from PYR (mode at 365 cm$^{-1}$) units is found to increase monotonically with x, displaying a plateau in the 25% < x < 28% range, and then to increasing further as x approaches 40%. Scattering from QT units (333 cm$^{-1}$ mode) reveals a broad maximum in the RW, 23% < x < 29% range (Figure 13). Furthermore, scattering from As=S stretch of QT units (490 cm$^{-1}$) also shows a broad maximum centered near x = 25%. The DFT calculations reveal that the Raman cross sections of PYR (360 cm$^{-1}$) and QT units (335 cm$^{-1}$) are 31.3 and 60.7 units respectively. Given these data, we are lead to believe that population of the QT units maximize near the onset ( x = 22%) of the RW. As expected the concentration of PYR units maximize near x ~ 40%. Several observations can be made concerning the 490 cm$^{-1}$ mode. (i) the mode is observed for glass compositions all the way up to x = 40% (see inset of Figure 12 (ii) Mori et al.[71], have examined IR transmission experiments from these glasses and find that the 490 cm$^{-1}$ mode to be present in glasses all the way up to x = 40%. Findings (i)-(ii) would appear to rule out identification of the 490 cm$^{-1}$ mode as either due to $S_n$ chains or $S_8$ rings. Integrated scattering from the pair of closely spaced modes near 467 and 473 cm$^{-1}$ due to $S_n$ chains and $S_8$ rings is found to monotonically decrease as x increases to 35% and to nearly vanish as x increases to 40%. Finally, mode frequencies of both PYR and QT units steadily *red shift* as x increases to 40%. Infrared reflectance and Raman scattering results of Lucovsky[74] on the stoichiometric glass at x = 40% ($As_2S_3$) have placed the symmetric and asymmetric stretch of PYR units at 340 cm$^{-1}$ and 309 cm$^{-1}$ respectively. These results confirm the steady red shift of these vibrations as x increases from 8% to 40%.

We had noted earlier the anomalously low $T_g$s ( Figure 5) and high molar volumes of As-S glasses (Figure 6) when compared to those of As-Se glasses . These data strongly suggest that As-S glasses are examples of networks that are *partially* polymerized, to be contrasted to the case of As-Se glasses that are examples of nearly *fully* polymerized networks. Our speculation is that QT and PYR units along with some $S_n$ chains form the backbone of As-S glasses. Some fraction of $S_n$ chain fragments demix from the backbone and contribute to free volume in the glasses reflected in their high molar volumes (Figure 6). The demixed $S_n$ chain fragments, like $S_8$ rings, couple to the backbone by weaker van der Waals forces instead of covalent ones. The demixed $S_n$ chain fraction steadily decreases as x exceeds 35%. The loss in elastic stiffness of the backbone can then be traced to an "effective" reduced connectivity due to presence of both weak van der Waals and strong covalent forces, and contributes to the a red-shift of Raman vibrational modes. These ideas on partial demixing of As-S glasses on a nanoscale, we will revisit when we discuss results on P-S glasses next.

*5.2 Binary $P_xS_{1-x}$ glasses*

A summary of Raman scattering in these glasses [51] at several compositions appears in Figure 14. In the range of glass formation, 5% < x < 22%, scattering is dominated by three broad modes near 377 cm$^{-1}$, 416 cm$^{-1}$ and 470 cm$^{-1}$. In addition, there are two closely spaced modes in the 700 cm$^{-1}$ region. These Raman results are quite similar to the earlier results of Koudelka et al. [75]. For comparison we have also included in Figure 14, Raman scattering of orthorhombic-S and of crystalline $P_4S_{10}$. Both these reference crystals are molecular solids composed of $S_8$ rings, and of $P_4S_{10}$ cages. A $P_4S_{10}$ cage may be visualized as made of 4 QT units. Based on first principles cluster calculations[72], we assign the three modes as follows; the 377 cm$^{-1}$ mode represents a symmetric stretch of QT units, the 416 cm$^{-1}$ mode a symmetric stretch of PYR units while the 470 cm$^{-1}$ mode a stretch of S-chains. The doublet observed near 700 cm$^{-1}$ we assign to P=S stretching vibration of the QT units. The present assignments are in harmony with earlier Raman[75] and NMR results[76]. At x > 20%, P atoms rapidly leach out from the backbone to form $P_4S_{10}$ cages, leading to the doublet feature near 700 cm$^{-1}$ to become asymmetric. Once bulk glass formation ceases at x > 22%, the Raman lineshapes are dominated by modes of $P_4S_{10}$ monomers and display several sharp features in the 100- 300 cm$^{-1}$ range (Figure 14) suggesting formation of a nano- or micro-crystalline phase. In orthorhombic S, one observes a sharp mode near 475 cm$^{-1}$ ascribed to presence of $S_8$ rings. In glasses, corresponding $S_8$ ring modes are also present but with a difference, the linewidth of the modes are noticeably broader suggesting that monomers do not condense into a crystalline phase but are interspersed in the glass network as isolated molecules. The absence of the α to β transition near 109°C in m-DSC scans of the present glasses, corroborates that $S_8$ rings do not condense to form the α phase in these binary glasses.

The normalized scattering strength ratio of the various modes observed in Raman scattering of P-S glasses appears in Figure 15, and it can serve as a basis to understand the structure evolution of these glasses. In the 5% < x < 12% range, $S_8$ ring fraction rapidly declines and the $S_n$ chain fraction rapidly grows as the P content of glasses is increased. Also in this range, the PYR and QT units increase in concentration. Perhaps the most striking features of these data is that in the RW, we find both QT (377 cm$^{-1}$) and PYR (416 cm$^{-1}$) units show a global maximum of concentration. These building blocks undoubtedly define the backbone. At x > 19%, $P_4S_{10}$ molecules comprising a nanocrystalline phase rapidly increase in concentration. The narrow glass forming range, 5% < x < 25%, in the P-S binary glasses is bounded by nanoscale phase separation effects, with $S_8$ units segregating on the low side and $P_4S_{10}$ cages on the high side of the glass forming range.

What can we say about the $S_n$ chain fragments? Are they part of the glass backbone or are they decoupled from it? Guidance on the issue comes directly from the $T_g$ variation of these glasses (Figure 4) and the location of the RW in r-space (Figures 8 and 16). At the low end of the RW and particularly in the 14% < x < 16% range, $T_g$ scaling between the P-Se and P-S glasses is quite low, and $S_n$ chains would appear to be *decoupled* from the backbone. In the RW, if we assume that only PYR and QT units comprise the backbone and that all $S_n$ chains are decoupled from these units, then we can predict the stoichiometry of the backbone. Taking the scattering strength of the PYR and QT units from the Raman scattering data (Figure 15) and normalizing these data to the known Raman cross-sections from the DFT calculations, we find that the concentrations of PYR and QT units in the RW to be about the same. Under the circumstance, stoichiometry of the glass backbone becomes $PS_{1.5} + PS_{2.5}$ or $P_2S_4$ or $P_{1/3}S_{2/3}$, which is identical to the stoichiometry of the RW centroid found in corresponding selenide glasses, $P_xSe_{1-x}$, x = 33.3. These data suggest that the large shift of the RW in P-S glasses to low *r* ( = 2.15) relative to the one in P-Se glasses (*r* = 2.35) is a structural effect- the *near complete decoupling* of $S_n$ chains from the backbone in P-S glasses, and the *complete coupling* of $Se_n$ chains in the backbone of P-Se glasses. At the higher end of the RW and particularly in the 16% < x < 19% range, some $S_n$ chains most likely become part of the backbone and this is reflected in an increased scaling of $T_g$. But as x exceeds 22%, $T_g$s reduce and glasses nanoscale phase separate and the glass forming tendency vanishes.

Returning to the case of As-S glasses, we had noted earlier (Figure 7) that the RW is shifted to the 22% < *r* < 28% range. Given the result above, we would like to suggest that binary As-S glasses represent the intermediate case, while P-S and P-Se glasses are respectively the two extremes of network connectivity. Specifically, $S_n$ chain fragments almost *completely decouple* from the backbone in P-S glasses, but they *partially decouple* in As-S glasses, while $Se_n$ chains *completely couple* to the network backbone in P-Se glasses. These aspects of glass structure related to the connectivity of their backbones apparently control the location of the RW in *r*-space. In this respect, RWs provide a new spectroscopy to probe connectivity of network glasses at intermediate and extended length scales.

## 6. Intermediate phases, local structural variance and network demixing

IPs for several families of chalcogenide glasses are summarized in Figure 16. Perusal of the data reveals several generic trends: (i) In the two binary group IV (Si, Ge) selenides, IPs reside[30, 31] in the $2.40 < r < 2.52$ range. (ii) On the other hand, in the two binary group V (P, As) Selenides, IPs shift to a lower connectivity[26, 56], $2.28 < r < 2.40$. (iii) Furthermore, ternary selenide glasses containing equal concentrations of the group IV and group V elements, IPs are found to encompass ranges of corresponding binary glasses. Thus, for example the IP in $Ge_xAs_xSe_{1-2x}$ glass system extends in the $2.28 < r < 2.48$ range[39], which almost covers the IP in $As_xSe_{1-x}$ glasses, $2.28 < r < 2.38$, and in binary $Ge_xSe_{1-x}$ glasses, $2.40 < r < 2.52$. (iv) In *ternary* sulfides[42, 43] IPs are centered in the same region of $r$-space as their selenide counterparts, but their widths, in general, are somewhat narrower than in corresponding selenides. (v) Finally, IPs in binary(P-S, As-S) sulfides reveal anomalous features; their centroids are shifted[9, 51] to much lower values in $r$, and their widths are significantly narrower than in corresponding selenides. We will now address questions such as, how are we to understand these data and in particular what aspects of glass structure control these IPs?

### 6.1. Binary group IV- Selenides

There is broad recognition that glass compositions in the IPs are composed of isostatic local structures which form the building blocks of the rigid but stress-free network[38, 44]. In the group IV selenides, the two isostatic local structures include CS $GeSe_4$ tetrahedra (chemical stoichiometry $r = 2.40$) and ES $GeSe_2$ (chemical stoichiometry $r = 2.67$) as shown in Figure 17. Long chains of ES tetrahedra become optimally constrained ($n_c = 3$) because in the planar edge-sharing $Ge_2Se_2$ contacts a bond angle constraint becomes redundant, and lowers the count of constraints per atom ($n_c$) from 3.67 to 3.0. Raman scattering in binary $Ge_xSe_{1-x}$ glasses shows that CS tetrahedra are the majority local structures at the onset of the IP at $x = 20\%$ or $r = 2.40$. ES $GeSe_2$ tetrahedra grow in concentration with increasing x and represent 31.4% of the scattering strength as x increases to 1/3 ($GeSe_2$). It is reasonable to assume that a network composed of varying proportions of these local structures contributes to the observed IP spanning the $2.40 < x < 2.52$ range (Figure 16).

Why does the IP extend to $r = 2.52$ only and not to $r = 2.67$? A possible reason could be that as $r > 2.40$,

CS GeSe$_4$ tetrahedra steadily become Ge-richer as the dimeric Se-Se bridges between Ge atoms are replaced by Se bridges. Near $r \sim 2.50$, on an average CS tetrahedra possess a GeSe$_3$ stoichiometry, which would require each Ge atom to have 2 Se bridges and 2 dimerized Se bridges, i.e., Ge (Se$_{1/2}$)$_2$Se$_2$. A count of constraints for such a Ge-richer CS tetrahedra yields n$_c$ = 3.25, i.e., such tetrahedra are mildly stressed rigid (n$_c$ > 3). In a mean-field sense, one can begin to see that the IP could not extend much beyond $r \sim$ 2.50 even though ES isostatic chains continue to be available till $r$ = 2.67. Simulations of Ge-Se glasses using a first principles density functional code FIREBALL by Inam et al.[49] confirm that CS GeSe$_4$ tetrahedra show a global maximum in the IP region. A mix of these two isostatic local structures provides the local structural variability to form IPs in the two group IV chalcogenides. That view is corroborated by the SICA calculations of Micoulaut and Phillips[45]. The concentration ratio of these two isostatic local structures as a function of glass composition can serve as an important check on realistic structural models[48] of IPs once these are constructed in large scale numerical experiments in future.

### 6.2. Binary group V Selenides

Binary P-Se glasses are composed of three types of local structures[56], and two of these are isostatic and include pyramidal( P(Se$_{1/2}$)$_3$ ) and a quasi-tetrahedral(Se = P(Se$_{1/2}$)$_3$ ) as illustrated in Figure 17. The third unit contains P-P bonds in a polymeric ETY like chains, which already manifest near x > 20%, and are mildly stressed-rigid (n$_c$ = 3.25). The chemical stoichiometry of quasi-tetrahedral (QT) units is $r$ = 2.29, and of pyramidal (PYR) units, $r$ = 2.40 (Figure 16). In P$_x$Se$_{1-x}$ glasses, one thus expects QT units to be populated predominantly at the start (x = 29%, or $r$ = 2.29), while the PYR units at the end (x = 40% or $r$ = 2.40) of the IP. $^{31}$P NMR experiments yield populations of these units (Figure 9) as a function of glass composition[56, 57, 66], and one finds the concentration of QT units to show a broad global maximum centered near x = 25%, while PYR units to show a maximum near x = 35%. ETY-like structures of $r$ = 2.50 stoichiometry, as expected, show a maximum near x = 50% or $r$ = 2.50. Their presence in the backbone serves to nucleate stress, and it is not surprising that the IP terminates near x ~ 40%, when the population of these local structures overwhelm the other two isostatic local structures (Figure 9).

The IP in binary As-Se and P-Se glasses bear a close similarity to each other; these start at $r \sim$ 2.29 and extend to a composition of r ~ 2.38 (figure 16). These results suggest that the commonality must also extend to some of their local structures. Although it is generally accepted that PYR units are the local structures present in binary As-Se glasses, such is not the case with QT units. $^{75}$As NQR experiments on binary $As_xSe_{1-x}$ glasses as a function of glass composition x have been reported[77], although a structural interpretation of these data is at present less clear. Are the EFG parameters of As based PYR local structure different enough from those of QT ones to permit discriminating these two local structures from the observed nuclear quadrupole coupling distributions . Some guidance from first principles calculations of EFG parameters[72, 78] of these local structures could help in addressing the issue. Generally speaking, one expects EFG parameters for both these local structures to be similar, and to be small, given that As represents a case of a half-filled p-shell. We have already mentioned that the observed slope, $dT_g/dx$, of $As_xSe_{1-x}$ glasses at x < 10% is too small to support PYR units as the only local structures in these glasses (section 3.1.2). Recently, a molecular dynamic study of $As_xSe_{1-x}$ glasses[47], finds evidence of a large fraction of 4-fold coordinated As in a QT local structure at x < 0.40. Furthermore, FT-Raman data on these glasses has been recently analyzed[64] using guidance from first principles cluster calculations. As illustrated here for the case of binary As-S glasses, deconvolution of the observed Raman lineshapes of As-Se glasses in the 200 cm$^{-1}$ to 280 cm$^{-1}$ region also reveals modes of QT and PYR units . More significantly, the scattering strength of the symmetric stretch of QT units shows a global maximum in the IP[64]. These new results suggest that the IP observation in both As-Se and P-Se binary glasses can be reconciled in terms of the same two local structures.

*6.3. Group V binary sulfides.*

The IP in binary P-S binary glasses is rather anomalous both for its width and its centroid; the former is rather narrow ($\Delta r$ = 0.03) while the latter is shifted to a rather low value of *r* = 2.17(Figure 16). The structural origin of the centroid shift can be traced to a complete demixing of the $S_n$ chain fragments from a backbone that is largely composed of QT and PYR units. Under that circumstance one finds (section 5) that the chemical stoichiometry of the backbone is $P_{33.3}S_{66.6}$, the same as that of the IP centroid found in P-Se glasses. The concentration dependence of the QT and PYR units in P-S glasses reveal a maximum in the IP (Figure 9). These results are strikingly parallel to those found in corresponding binary P-Se glasses. The much narrower width of the IP in the binary P-S system (section 5) can be traced to the stability of $S_8$–rings and of $P_4S_{10}$ cages to form and segregate respectively at the low-end (x < 14%) and the high-end (x > 19%) of the IP . Thus, available thermal and spectroscopic data suggests that both the shift and the width of the IP in binary P-S glasses in relation to P-Se glasses can be reconciled in terms of differences in bonding chemistry of S from Se.

The IP in As-S glasses resides in the $2.22 < r < 2.29$ range, and we comment on it next. Although the width of the IP is comparable to the case of the corresponding selenides, the centroid of the IP in the sulfide glasses is shifted to lower $r$ in relation to the corresponding selenides. The IP centroid shift can be traced to a <u>partial decoupling</u> of $S_n$ chain fragments from the backbone. The case of As-S binary appears to be intermediate to that of P-S binary on one extreme, where a <u>complete decoupling</u> of chalcogen chains occurs to that of the As-Se binary on the other extreme where <u>complete mixing</u> of $Se_n$ chains occurs with the backbone. Raman scattering experiments on As-S[9] and P-S glasses[51] show that backbone structure of both these glasses is composed of PYR and QT units with their concentrations maximizing in the IP. These data are persuasive in suggesting that in group V chalcogenides, the two isostatic local structures of relevance in the IP include the PYR and QT units.

The stoichiometric glasses, $As_2S_3$[19] and $As_2Se_3$[79], are important optoelectronic materials, and their physical properties are widely understood in terms of glass structures that are closely similar to their crystalline counterparts composed of fully polymerized networks of PYR units exclusively. If this were the case both these stoichiometric glasses would be self-organized and these compositions would be part of the IP. We can see from Figure 16 that this is nearly the case for $As_2Se_3$ glass, a composition that sits close to the upper end of the IP in binary As-Se glasses. On the other hand, in $As_xS_{1-x}$ glasses the composition, x = 40% or $r = 2.40$ is far from being in the IP. Secondly the $T_g$ of $As_2S_3$ glass should have been closer to 250°C rather than 190°C if the sulfide glass were to form a fully polymerized network as the selenide (section V). These data suggest that $As_2S_3$ glass is neither fully polymerized nor chemically ordered[60] as its crystalline counterpart. An intrinsic feature of the glass structure is the presence of substantial free volume (Figure 5), and such a network packing effect is likely to determine in a significant way how pair-producing light interacts with the material to influence optical non-linearity[80] and photodarkening[81] properties.

*6.4. Ternary Group IV and V selenide- and sulfide- glasses*

Ge as an additive in base P-S glasses brings about some rather remarkable changes in glass network morphology. Decoupled $S_n$ chain fragments present in binary P-S glasses apparently readily alloy with Ge additive, and homogeneous ternary $Ge_xP_xS_{1-2x}$ glasses[43] can be realized over a fairly wide range of composition x (Figure 3). In the ternary, presence of chemical disorder serves to largely suppress nanoscale phase separation effects characteristic of binary glasses[12]. That view is corroborated by the chemical bond strength scaling of $T_g$s (Figure 3) observed in the ternary Ge-P-Se and Ge-P-S glasses, but not in the binary glass systems such As-Se and As-S ones (Figure 5).

The IP in ternary sulfides possesses features that have close parallels to those in corresponding selenides (Figure 16). The IP widths in the ternary sulfides (Ge-P-S, Ge-As-S) are narrower than those in corresponding selenides (Ge-P-Se, Ge-As-Se), largely because of (a) a tendency of sulfur-rich glasses to demix with the stable $S_8$ rings decoupling from the network backbone at the low end of the IP, and (b) tendency of binary sulfides such as $As_4S_4$, $As_4S_3$, $P_4S_{10}$, $P_4S_7$ to form monomers and readily decouple from the backbone at the high end of the IP (Figure 16). An extensive review of the material properties of ternary $Ge_xAs_xS_{1-2x}$ glasses was recently provided by Kincl and Tichy[69].

In the present approach the outstanding features, (i) –(v), of IPs listed at the beginning of this section can be reconciled in a natural fashion in terms of the four isostatic local structures shown in Figure 17. Presence of these four local structures in the IPs is supported by available calorimetric, Raman scattering and constraint counting algorithms. These aspects of local structure would appear to provide the structural variance contributing to widths of these phases. Recently Lucovsky et al.[82] and independently Sartbaeva et al.[83] have suggested other avenues to build structural variance in chalcogenides to understand IPs, and the interested reader is steered towards their work. Understanding the local and intermediate range structure of IPs may also hold the key to reconciling their extraordinary functionalities. [84]

**Concluding remarks**

Variation of $T_g$ and of non-reversing enthalpy ($\Delta H_{nr}$) of $T_g$ as a function of chemical composition in binary and ternary chalcogenide glasses in conjunction with Raman scattering have provided unprecedented details on network connectivity, molecular structure and elastic behavior of their backbones. Tools of constraint theory[32], stochastic agglomeration theory[52] and numerical simulations of the vibrational behavior of covalent networks[33, 34, 44] and characteristic clusters[72] have proved to be invaluable in decoding results of these experiments. Discovery of reversibility windows in these glasses[29] that coincide with Raman elastic thresholds has facilitated identification of regions of compositions that self-organize to form Intermediate Phases (IPs). Widths and centroids of IPs in chalcogenide glasses can be broadly reconciled in terms of four *isotatic* local structures (Figure 17), two of these units (PYR($r$ = 2.40), QT($r$ = 2.28)) are populated in binary group V-Sulfides and selenides, while the other two (CS ($r$ = 2.40) and ES tetrahedra ($r$ = 2.67)) in group IV–Sulfides and selenides. In ternary alloys containing equal concentrations of group IV and group V elements with Selenium or Sulfur, all four of these local structures are known to exist in the backbone, permitting many more possibilities for networks to reconfigure and self-organize and form IPs. Since stoichiometries of these 4 local structures span a wide range of connectivity from $r$ = 2.28 to $r$ = 2.67, IPs in ternary glasses are wider than in binary glasses. These aspects of structure also explain in a natural fashion, why IPs in binary group V selenides are shifted to lower $r$ in relation to those in binary group IV- selenides. Of special interest are binary group V-sulfides in which IP centroids are shifted to even lower $r$ in relation to corresponding selenides. The feature is thought to result from demixing of $S_n$ chain fragments from backbones in the former binaries but the complete mixing of $Se_n$ chains with the backbone in the latter ones. The demixing of excess S is partial in nature in the As-S binary, but it is almost total in the P-S binary.

**Acknowlegements**

It is a pleasure to acknowledge discussions with Professor Bernard Goodman, Professor Jim Phillips, Professor Matthieu Micoulaut, Professor Alan Jackson, Deassy Novita and Professor Daniel Georgiev during the course of this work. This work is supported by NSF grant DMR-04-56472.

*Present address: Department of Materials Science and Engineering, Johns Hopkins University, Baltimore, MD 21210.

**Figure Captions**

**Figure 1**. $T_g$s in binary $Ge_xSe_{1-x}$ (●) and $Si_xSe_{1-x}$ glasses (○) [30, 53]. As low x ( < 10%) the linear variation yields a slope $dT_g/dx$ that is in excellent agreement with predictions[52] of Stochastic Agglomeration Theory (SAT) if Si and Ge are taken to be 4-fold coordinated.

**Figure 2**. Composition dependence of $T_g$ in binary $P_xSe_{1-x}$ [41] and in $As_xSe_{1-x}$ [26]. glasses as a function of group V atom concentration. The mean coordination number <r> is estimated as 2 + x. At low x (< 10%) the observed slope of 4.1°C/at % of P or As departs significantly from the SAT predicted slope of 7.7 C/at % of As or P, if the group V atom is taken to be 3-fold coordinated only.

**Figure 3**. Composition dependence of $T_g$ in ternary $Ge_xP_xSe_{1-2x}$ [5] and in $Ge_xP_xS_{1-2x}$ [43] glasses as a function of x. Group V atom concentration. The mean coordination number <r> is estimated as 2 +3x.

**Figure 4**. Composition dependence of $T_g$ in binary $As_xSe_{1-x}$ [26] and in $As_xS_{1-x}$ [9] glasses as a function of x. The (□) data gives the $T_\lambda$ transition for the sulfide glasses. The broken line plot gives the bond-strength scaled $T_g$ of $As_xS_{1-x}$ glasses. See text for details.

**Figure 5**. Molar volumes of $As_xSe_{1-x}$ (■) [26] and $As_xS_{1-x}$ (●) [9] glasses as a function of As fraction x. Molar volume of c-$As_2S_3$(○) and c-$As_2Se_3$(□) taken from ref. [62] and ref [61], are also shown. The broken like is the predicted molar volumes of $As_xS_{1-x}$ glasses normalizing the $As_xSe_{1-x}$ glass data for the reduced size of S in relation to Se.

**Figure 6**. Non-reversing enthalpy ($\Delta H_{nr}$) at $T_g$ in $As_xS_{1-x}$ glasses [9] compared to the one in corresponding selenides, $As_xSe_{1-x}$ [26]. The reversibility window corresponds to the global minimum in $\Delta H_{nr}$ term and represents the Intermediate Phase as indicated.

**Figure 7**. Non-reversing enthalpy ($\Delta H_{nr}$) at $T_g$ in $P_xS_{1-x}$ glasses [51] compared to the one in corresponding selenides, $P_xSe_{1-x}$ [41].

**Figure 8**. Concentration of different local structures in binary $P_xSe_{1-x}$ glasses as a function of P content x deduced from $^{31}P$ NMR measurements. See [41].

**Figure 9**. Raman scattering in $P_xSe_{1-x}$ at x = 29% showing the various modes observed and their assignment. See text for details.

**Figure 10**. Modulated DSC scan of (a) pure S showing the α to β transition, liquidus and the sulfur polymerization $T_\lambda$ transition and of (b) a $As_{15}S_{85}$ bulk glass showing the $T_g$ (60°C) and $T_\lambda$(139°C) transitions. The $T_\lambda$ transition in the glass displays a precursive exotherm as seen in the non-reversing scan, a feature that would be difficult to ascertain from the total heat flow scan alone as in a traditional DSC measurement.

**Figure 11**. FT-Raman scattering from bulk $As_xS_{1-x}$ glasses at indicated As concentrations in percent on the left side of each scan. The x = 15% glass lineshape is deconvoluted in several modes, and come from $S_8$ rings $S_n$ chains, symmetric ($QT^{ss}$) and asymmetric ($QT^{as}$) stretch of QT, S=As$(S_{1/2})_3$ units, and an As=S stretch (490 cm$^{-1}$) of QT units, symmetric ($PYR^{ss}$) and asymmetric ($PYR^{as}$) stretch of PYR, As$(S_{1/2})_3$ units. At higher x ( > 40%), narrow modes of $As_4S_4$ Realgar molecules demix from network backbone. Note that the 490 cm$^{-1}$ mode is observed all the way up to x ~41%. See text for details.

**Figure 12**. Normalized scattering strength ratio of the $QT^{ss}$ mode (○) shows a broad maximum in the reversibility window range, PYR mode (●) progressively increases with x to show a maximum near x~ 40%, Sulfur chain and ring modes (■) steadily decrease with x to vanish near x ~40%, and the As=S stretch mode of QT units (Δ) show a broad maximum in the reversibility window. Matrix element effects are not considered in making the plot.

**Figure 13**. Raman scattering in binary $P_xS_{1-x}$ glasses at indicated P concentration in percent on the right, along with those of two reference materials, orthorhombic S and c-$P_4S_{10}$. In the composition range 14%< x < 19% where the reversibility window occurs, spectra are dominated by three modes, QT units, PYR units and S-chains. At x < 14%, modes of $S_8$ rings appear, while at x > 19% those of micro-crystalline $P_4S_{10}$ phase appear as glasses demix.

**Figure 14**. Normalized Raman scattering strengths of $S_8$ rings (◊) , $S_n$ chains ( Δ ) , PYR units (■) and QT units (●) , and amorphous $P_4S_{10}$ phase (▼) in binary $P_xS_{1-x}$ glasses as a function of P content x. This figure is adapted from[51].

**Figure 15**. Observed reversibility windows represent the calorimetric signature of Intermediate Phases in indicated binary and ternary chalcogenide glasses. In the bottom panel we show the connectivity of the 4 isostatic building blocks thought to provide the structural variance contributing to the IPs. See Figure 16 for a ball stick model of these local structures.

**Figure 16**. Local structure of Corner-sharing (CS) GeSe$_4$ tetrahedra (upper left), Edge-sharing GeSe2 tetrahedra (upper right), Quasi-tetrahedral (QT) AsS$_{5/2}$ ( lower left) and Pyramidal As$(S_{1/2})_3$ units. Each of these local structures is isostatic, i.e., the count of bond-stretching and bond-bending force constraints gives a count of 3 per atom. The stoichiometry of these local structures is indicated in terms of their connectivity *r*, and varies from 2.29 to 2.67. These structures comprise the elements of local structure contributing to the structural variance in the IP.

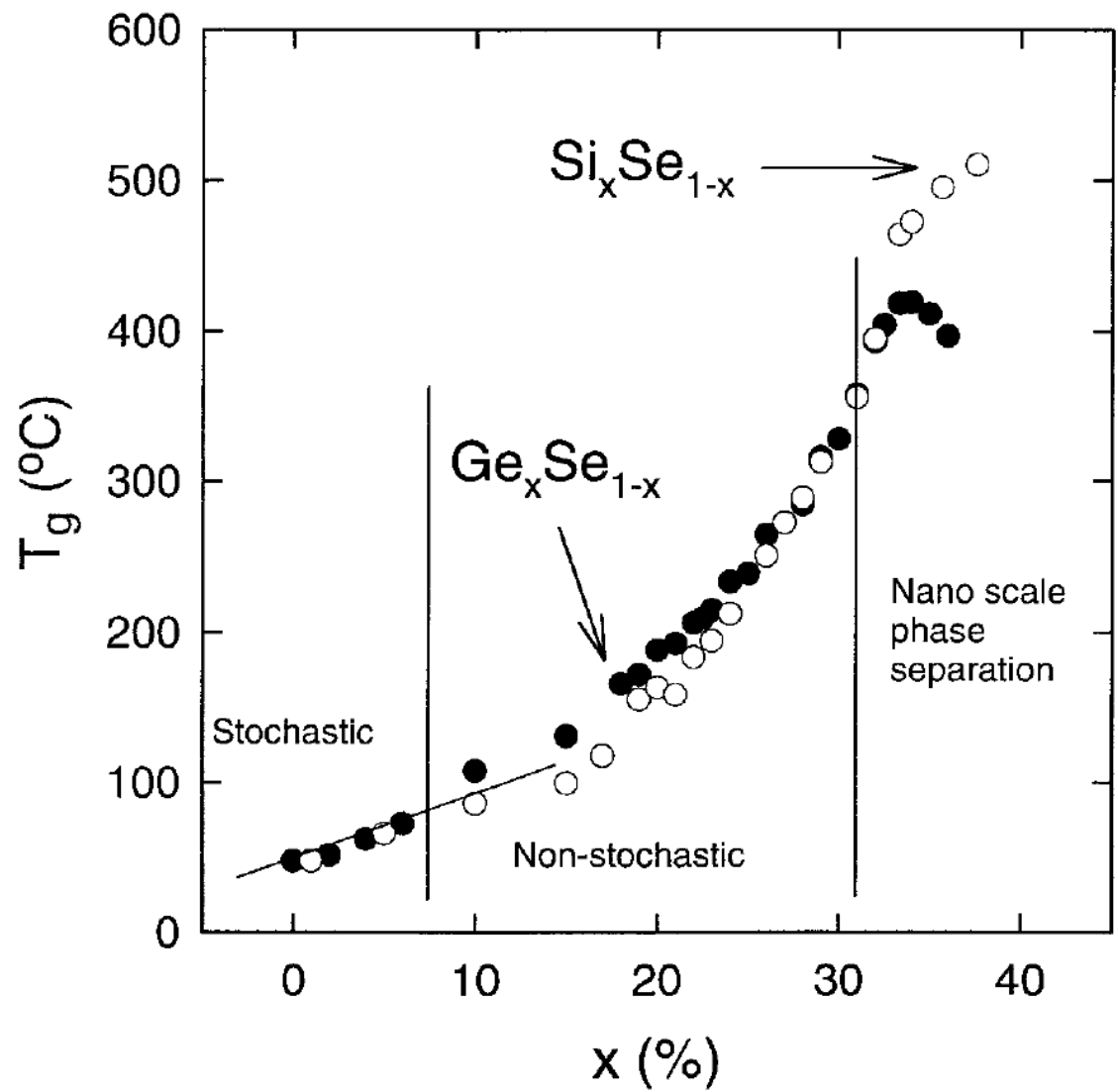

**Figure 1**

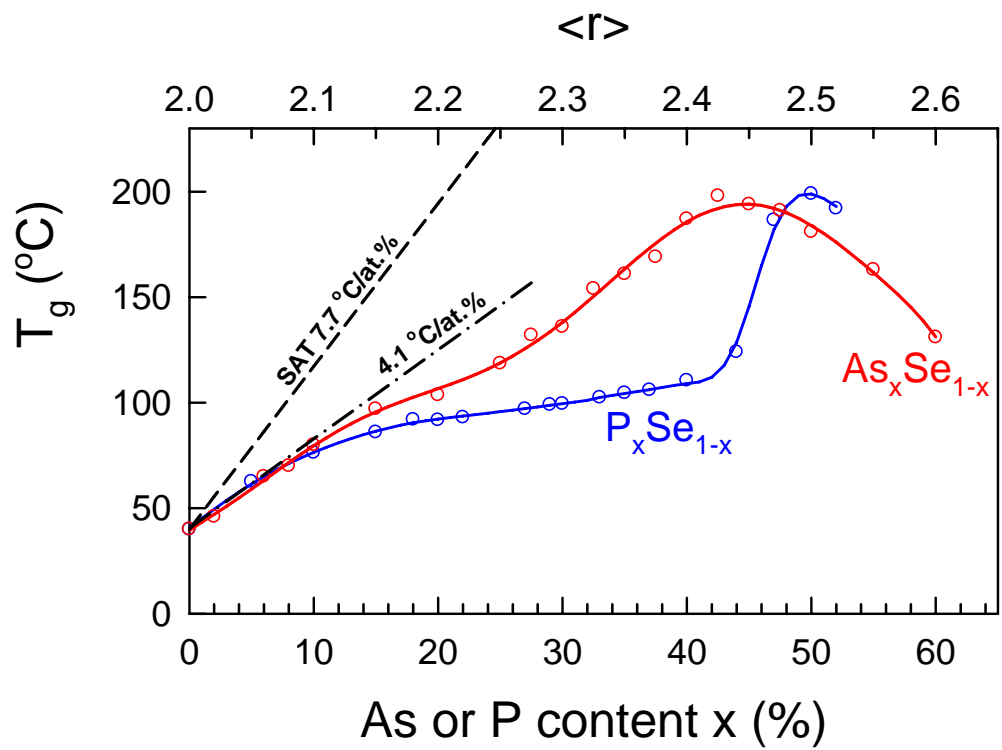

**Figure 2**

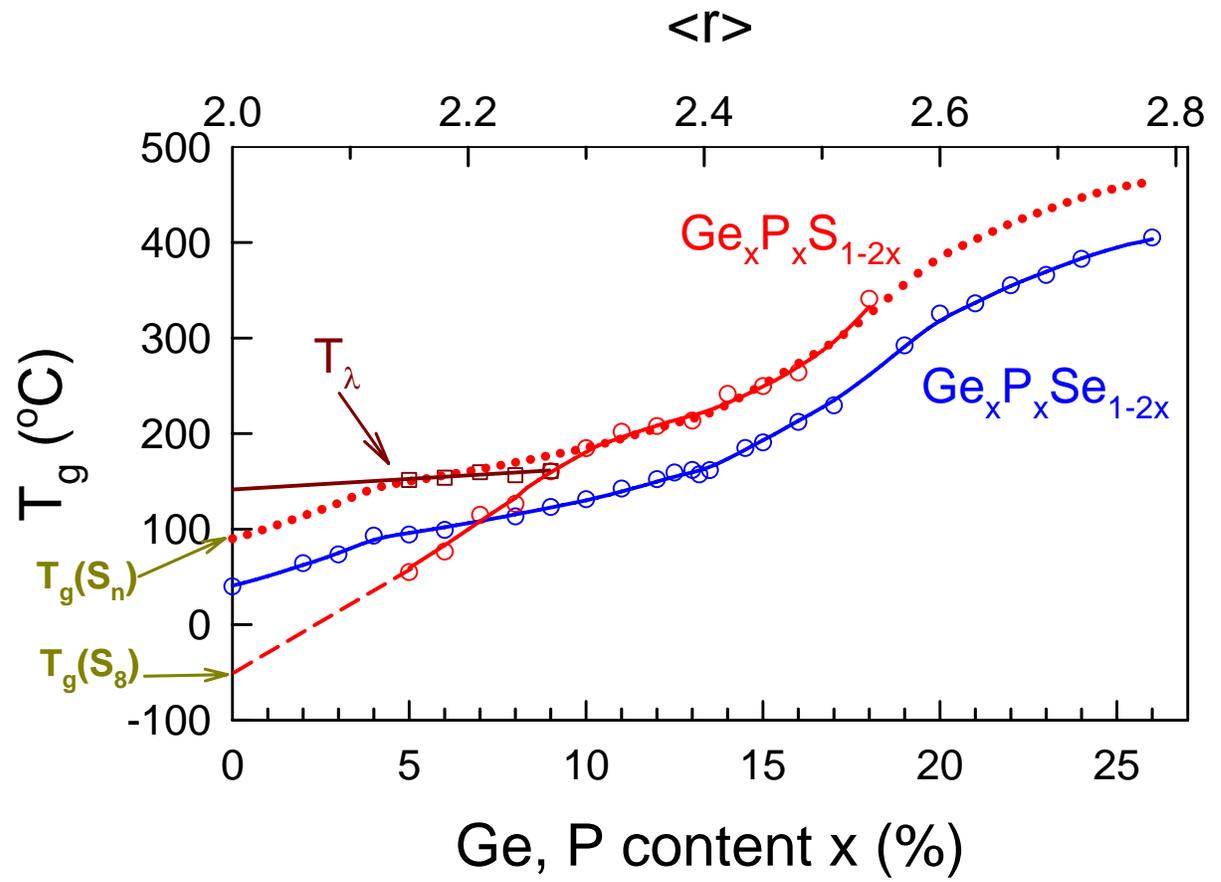

**Figure 3**

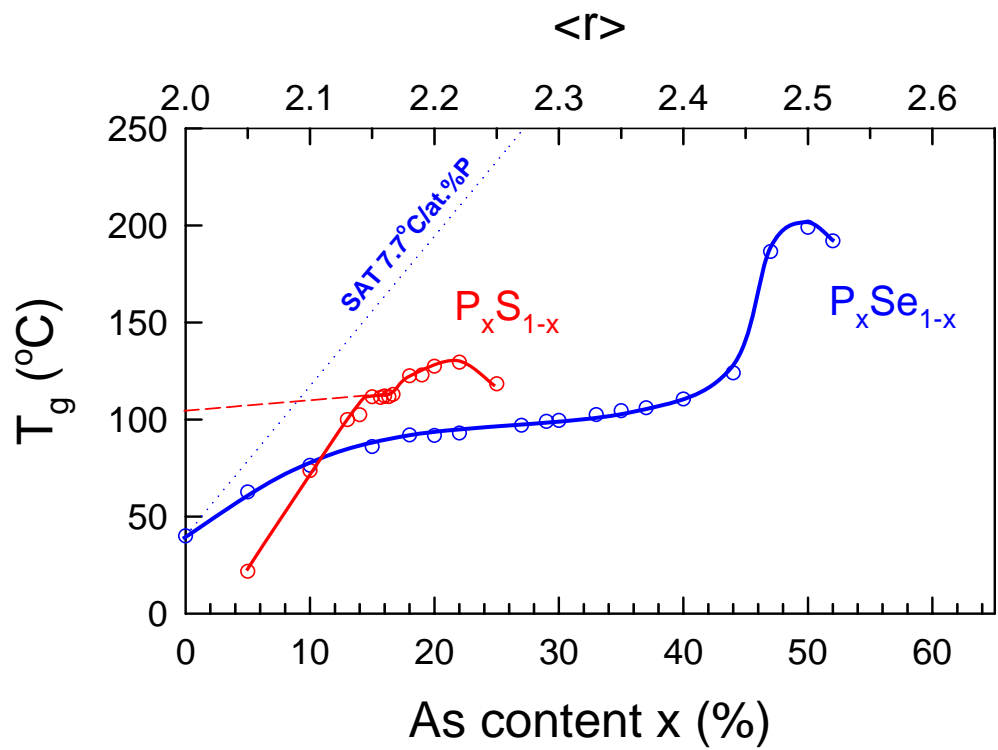

**Figure 4**

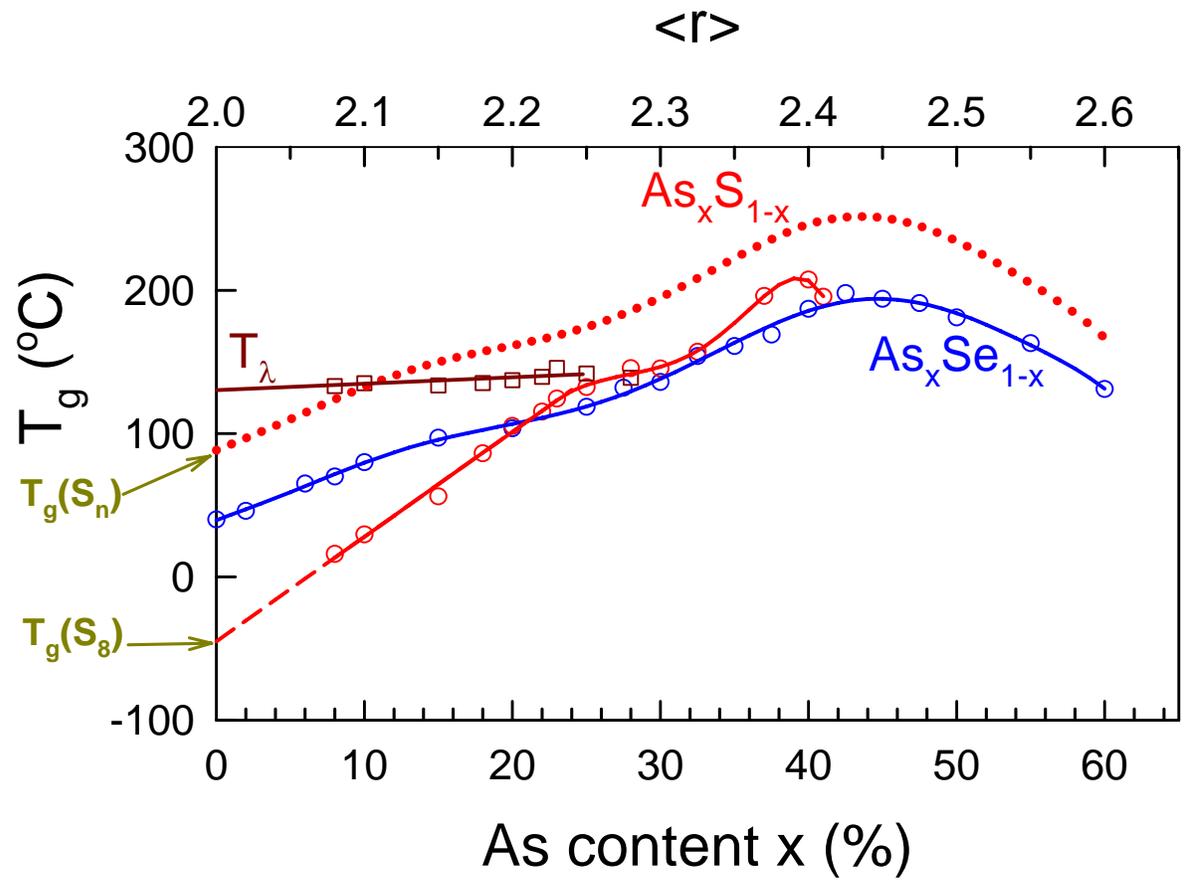

**Figure 5**

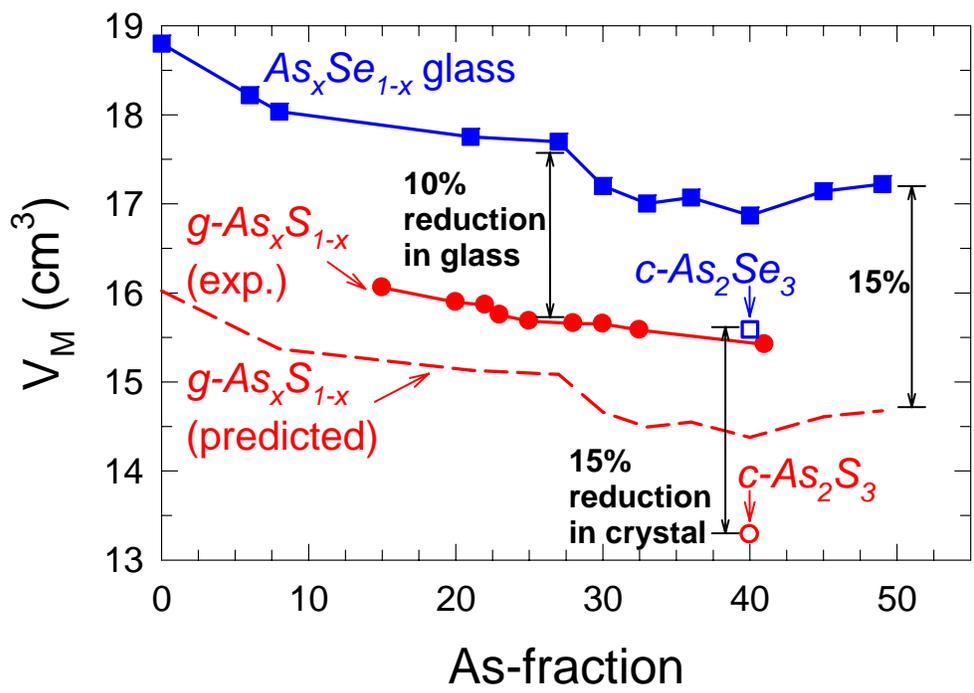

**Figure 6**

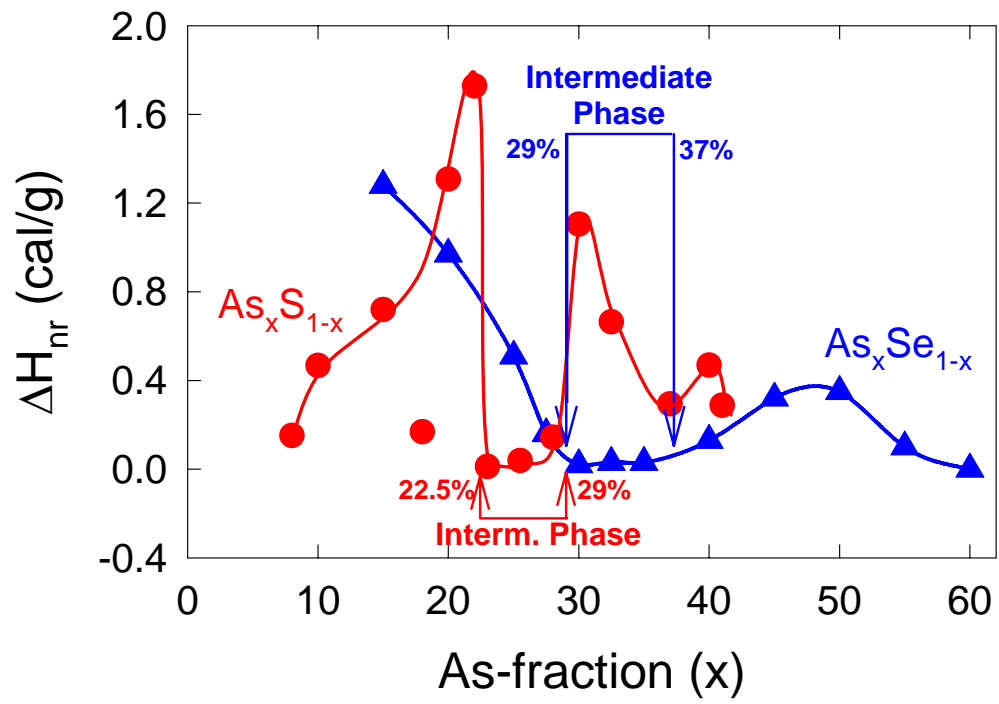

**Figure 7**

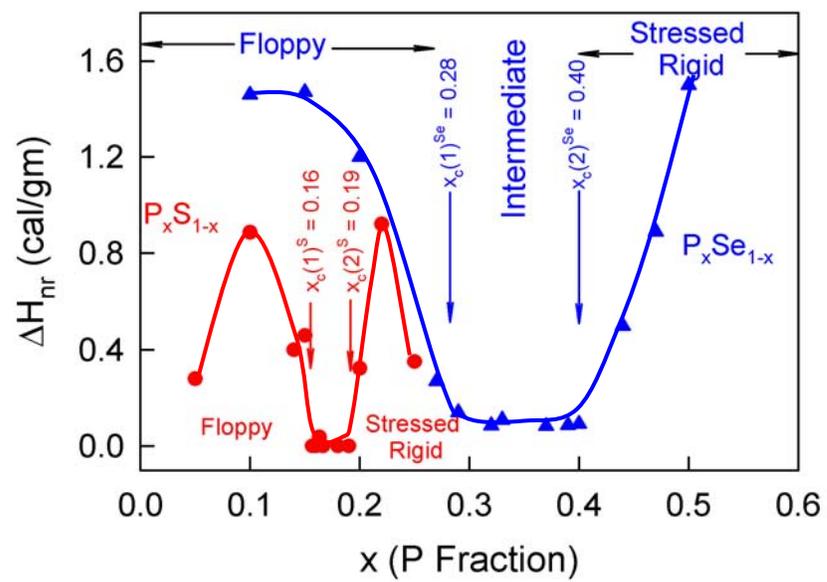

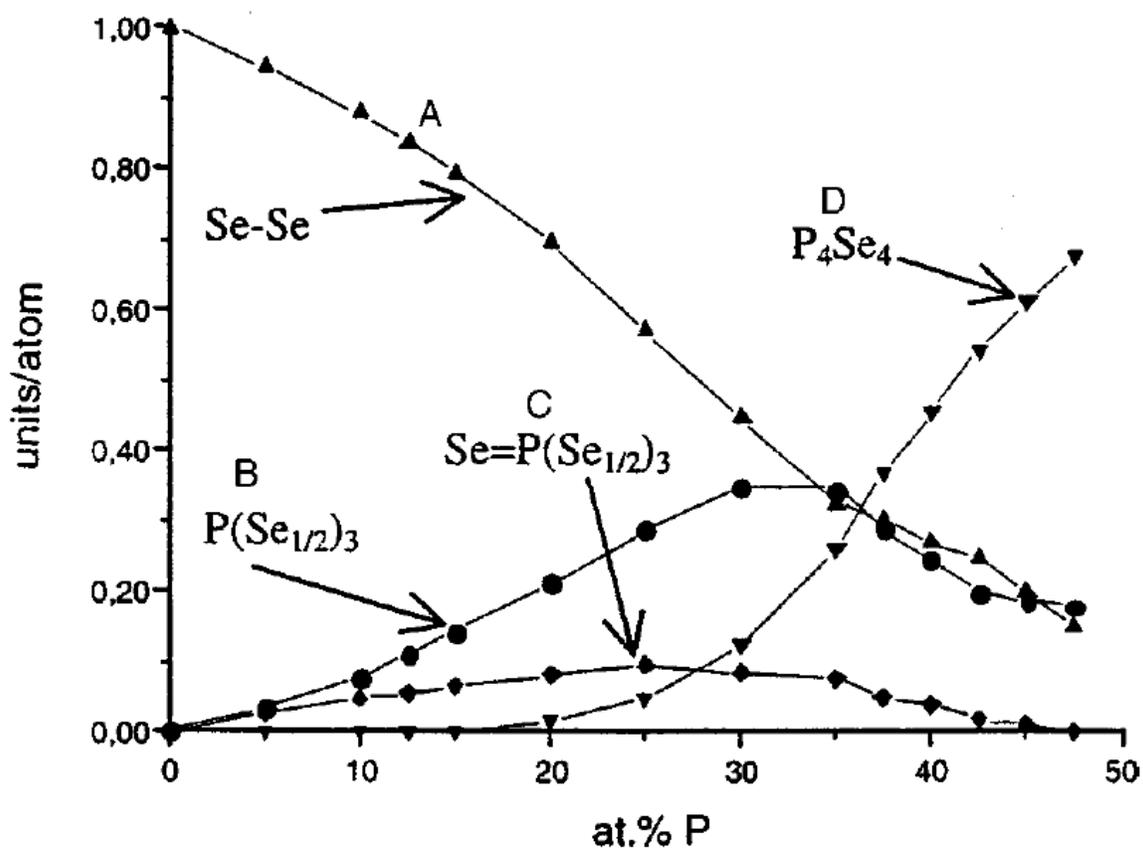

Figure 9

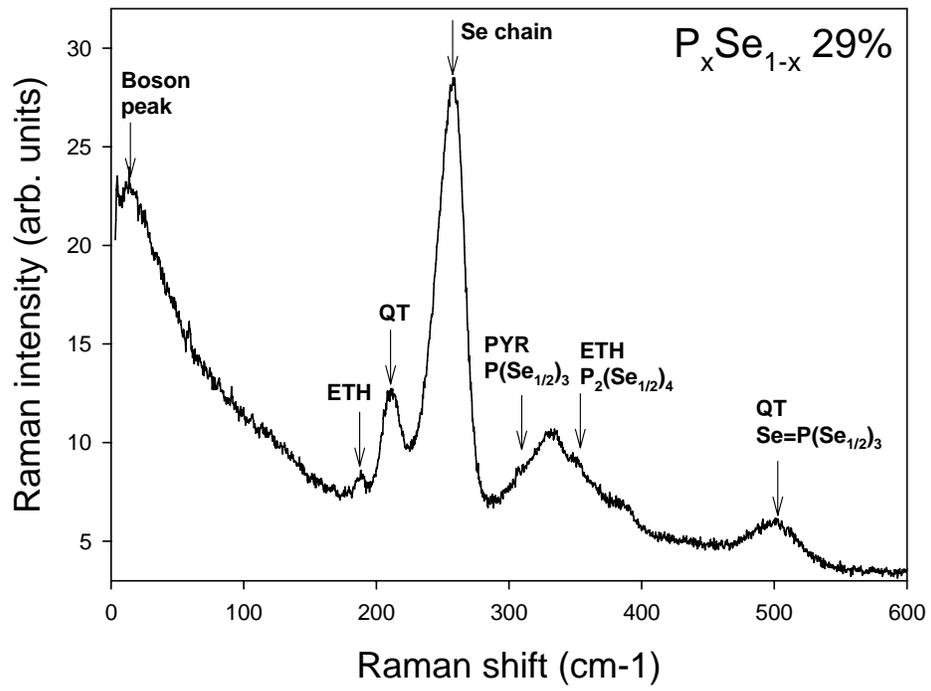

**Figure 10**

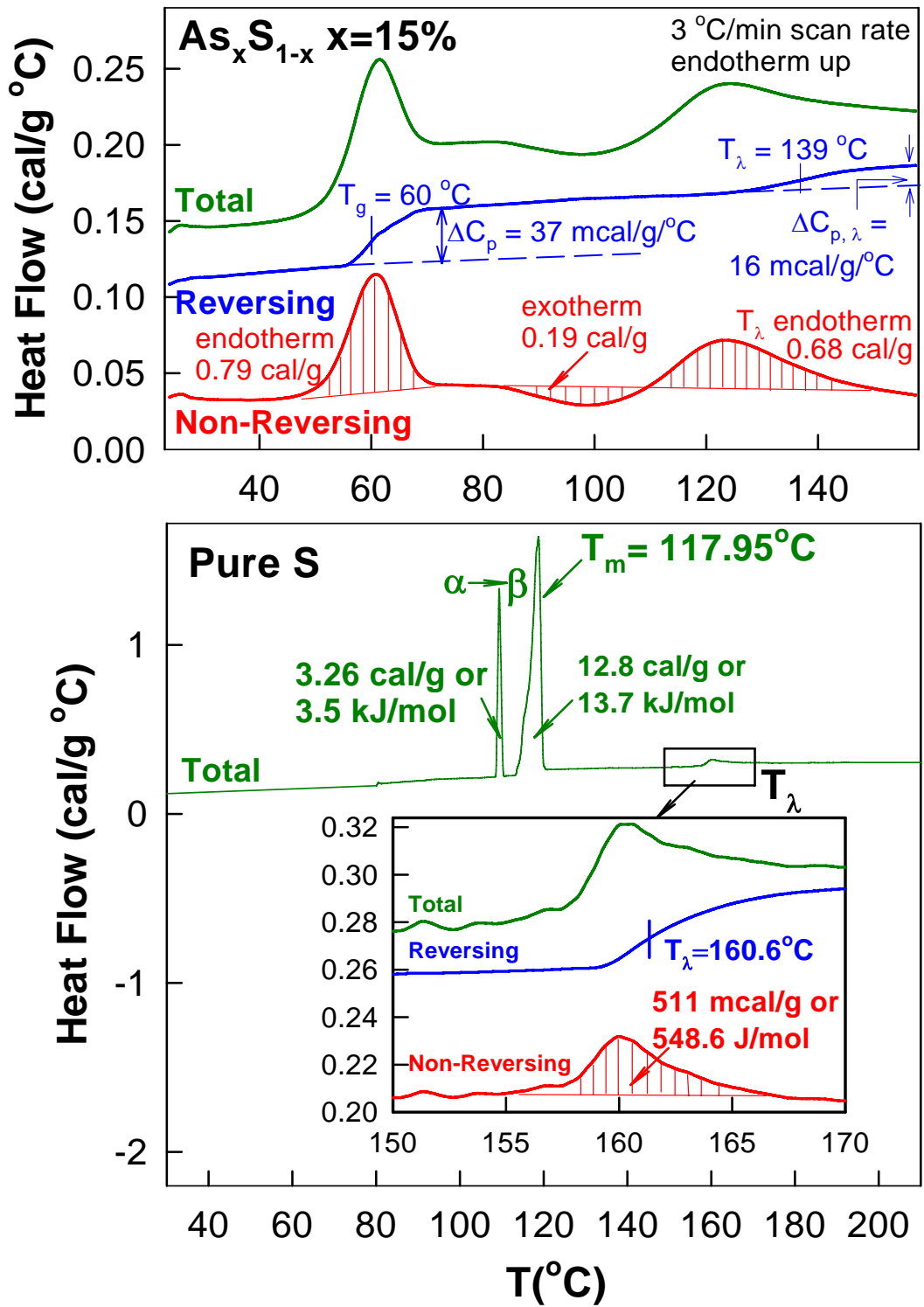

**Figure 11**

**Figure 12**

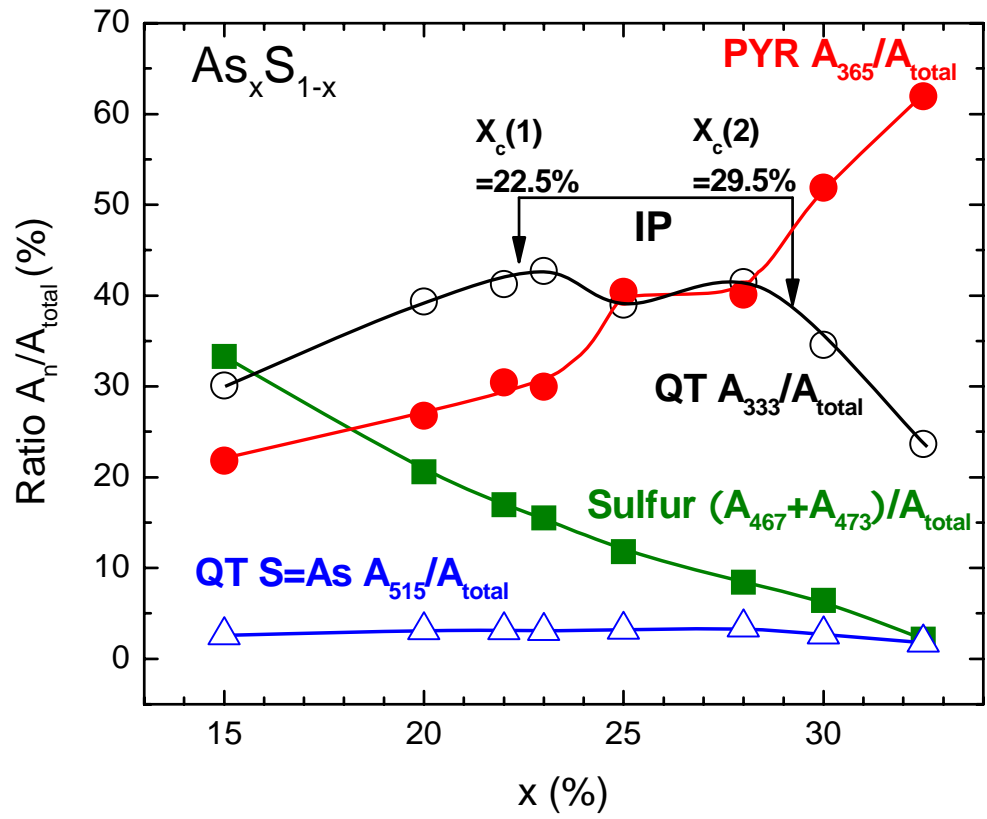

Figure 13

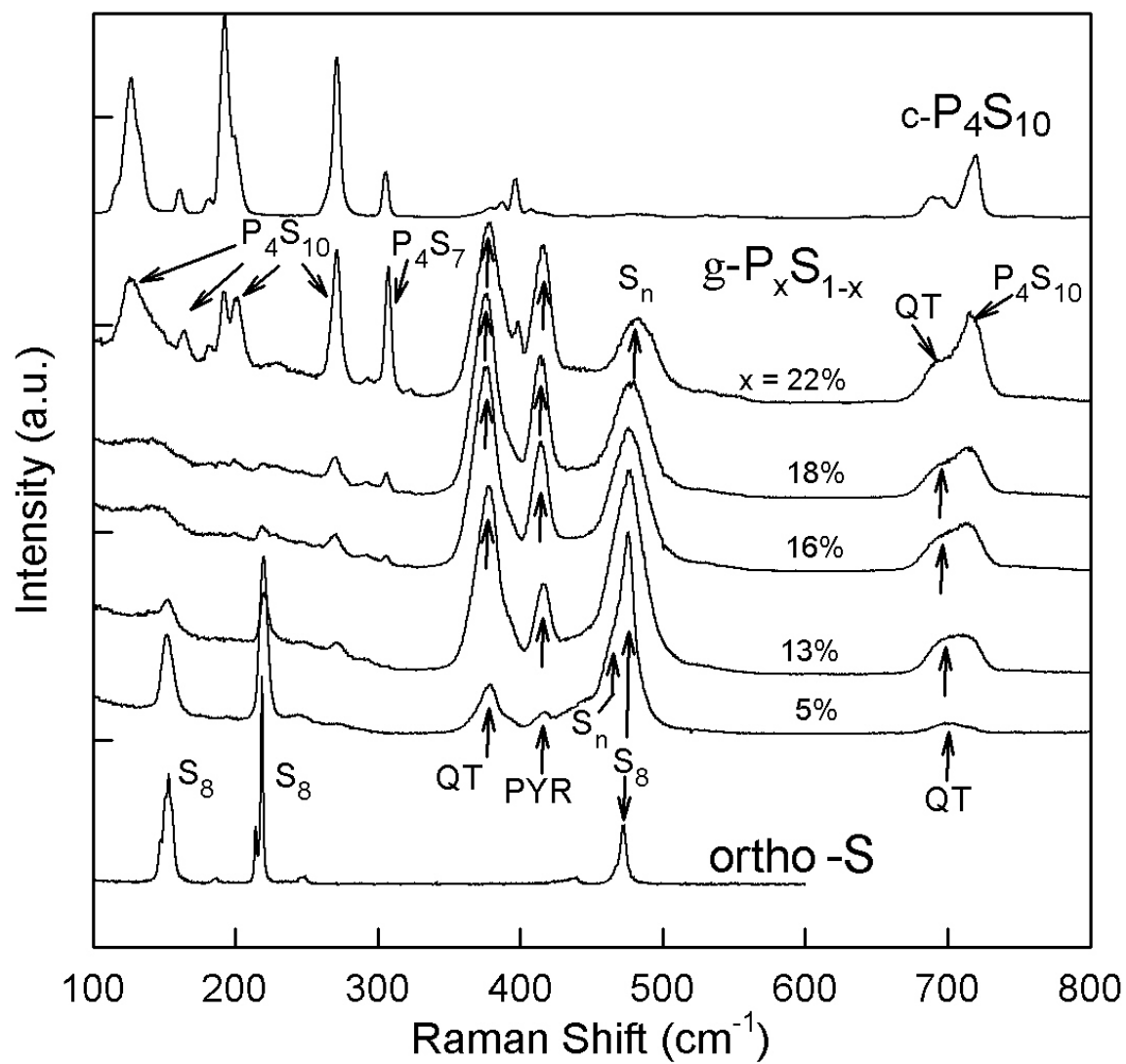

Figure 14

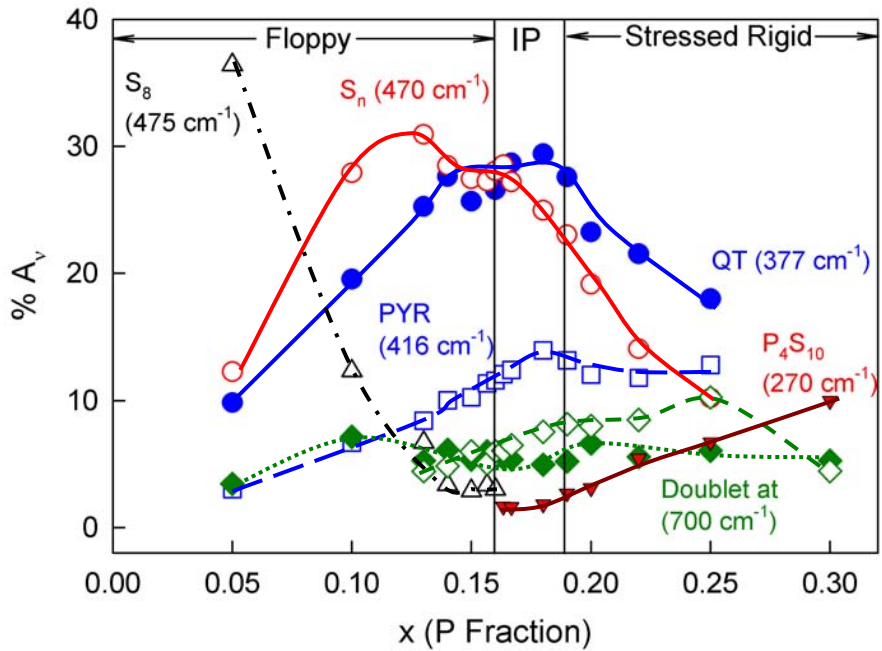

**Figure 15** Scattering strengths of various Raman modes observed in $P_xS_{1-x}$ glasses plotted as a function of Phosphorus concentration, x. The doublet feature at 700 cm$^{-1}$ is resolved into a mode at 700 cm$^{-1}$ (Filled diamonds) due to S=P stretch vibration in QT mode and other at 720 cm$^{-1}$ mode (Open Diamonds) due to S=P stretch vibration in $P_4S_{10}$ units. The $P_4S_{10}$ mode scattering strength (Inverted red triangles) increases monotonically starting from x = 19%.

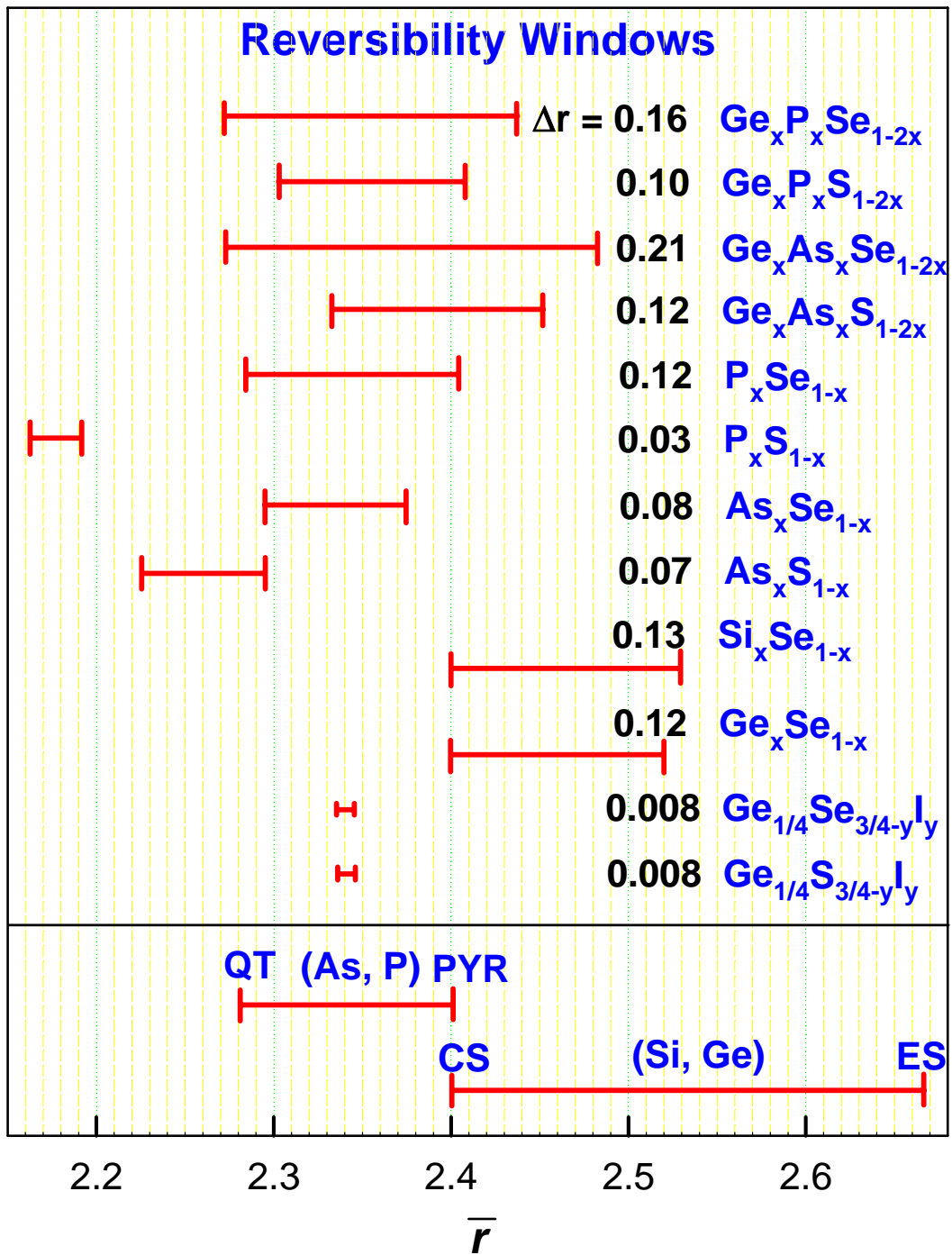

**Figure 16**

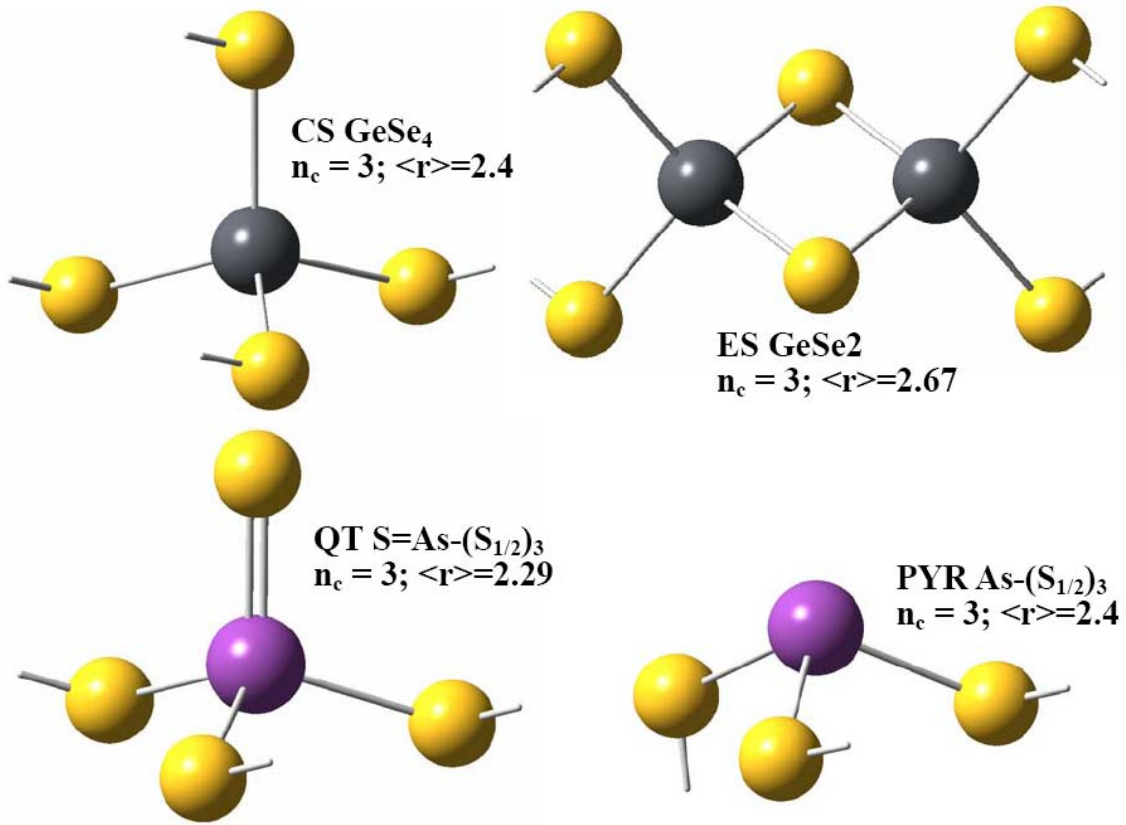

**Figure 17**